\documentclass[letterpaper]{article}
\usepackage{aaai2026}
\nocopyright % preprint: no AAAI copyright slug
\usepackage{times}                   % DO NOT CHANGE THIS
\usepackage{helvet}                  % DO NOT CHANGE THIS
\usepackage{courier}                 % DO NOT CHANGE THIS
\usepackage[hyphens]{url}            % DO NOT CHANGE THIS
\usepackage{graphicx}                % DO NOT CHANGE THIS
\usepackage{natbib}                  % DO NOT CHANGE THIS AND DO NOT ADD ANY OPTIONS TO IT
\usepackage{caption}                 % DO NOT CHANGE THIS AND DO NOT ADD ANY OPTIONS TO IT
\usepackage{amsmath}
\usepackage{amssymb}
\usepackage{bm}
\usepackage{booktabs}
\usepackage{tabularx}
\usepackage{multirow}
\usepackage{tikz}
\usetikzlibrary{arrows.meta, positioning}

\title{Stability of AI Governance Systems: A Coupled Dynamics Model
       of Public Trust and Social Disruptions}

\author{
    Jiaqi Lai,
    Hou Liang,
    Weihong Huang
}
\affiliations{
    Nanyang Technological University, Singapore
}

\begin{document}

\maketitle

% ============================================================
%  Abstract
% ============================================================
\begin{abstract}
AI systems are increasingly entrenched in public governance, yet
scholarship lacks formal tools to determine when deviations of
public trust in algorithmic institutions dissipate and when they
grow into collapse. Stability refers here to asymptotic recovery
from finite state perturbations under fixed structural parameters.
We address this gap by developing a
mathematical framework for institutional trust stability that couples
a Friedkin--Johnsen opinion dynamics process with a Hawkes-inspired
intensity process for AI controversies. Motivated by the
Computers-Are-Social-Actors literature and recent studies of trust in
large language models, this bidirectional coupling
reveals that governance stability depends on the structural
architecture of the information environment rather than absolute
trust levels. We derive an exact spectral stability criterion
delineating resilience from collapse, demonstrating how event
self-excitation and memory persistence systematically narrow the
stable parameter regime. Our structural analysis yields four
counterintuitive structural implications: high-trust systems can be structurally
fragile, low-trust environments can be structurally stable, dynamical
stability neither measures nor guarantees algorithmic fairness or
legitimacy, and network topology reshapes equilibrium heterogeneity
while its effect on spectral stability is uniformly bounded in an
explicit memory-dominated regime. Governance assessment should
therefore pair normative evaluation of harms and fairness with
structural analysis of recoverability, rather than treating either as
a proxy for the other.
\end{abstract}

% ============================================================
% ===== Section 1: Introduction =====
% ============================================================
\section{Introduction}
\label{sec:intro}

Artificial intelligence is no longer confined to research laboratories; it is increasingly embedded in the fabric of public governance. From algorithmic resource allocation in healthcare and welfare services to automated decision-making in criminal justice and urban planning, AI systems now mediate high-stakes choices that directly affect citizens' lives \citep{Levy2021PublicSector,Bright2025GenAIPublicSector}. Public trust is one important condition shaping the legitimacy and sustainability of these deployments \citep{Gillespie2023TrustAI,Edelman2023TrustBarometer}.

Yet public trust in AI-driven governance is not a passive quantity waiting to be measured---it is a dynamic, socially constructed state that co-evolves with the controversies surrounding deployment. When an AI system is perceived as producing unfair resource allocations or exhibiting systematic bias, the resulting public backlash need not remain isolated. Media coverage, social amplification, and networked attention can create self-reinforcing cascades \citep{Kasperson1988SARF,Zhao2015SEISMIC}. Trust, moreover, is asymmetrically fragile---far easier to destroy than to rebuild \citep{Slovic1993PerceivedRisk}---and trust--reliance dynamics are well documented at the individual level \citep{Lee2004Trust}. We argue that coupling trust to perceived controversy creates a distinct dynamical question: under a specified architecture, do perturbations decay or grow? Normative frameworks identify fairness, transparency, and accountability as governance requirements \citep{Jobin2019GlobalLandscape,Floridi2018AI4People}; the cited frameworks do not themselves provide a recovery criterion for this coupled trust--controversy process. Our model addresses that narrower question without treating stability as a substitute for those requirements.

A foundational modeling question is whether social-influence mechanisms developed for interpersonal settings remain useful first-order abstractions when the trustee is an AI system. Classical CASA research shows that people often respond socially to computers, while revisionist work documents medium-specific scripts and measurable differences from interpersonal trust \citep{Nass1994CASA,Gambino2020CASA}. Recent LLM experiments likewise show that anthropomorphic cues and uncertainty language affect perceived accuracy, trust, and reliance in context-dependent ways \citep{Cohn2024Anthropomorphism,Kim2024LLMUncertainty}. We therefore use \emph{social-update continuity} only as an explicit working hypothesis, not as a claim of substrate invariance or quantitative equivalence; Section~\ref{sec:scope} states its scope.

Meanwhile, the mathematical modeling community has developed powerful tools for studying related phenomena---opinion dynamics in social networks \citep{friedkin1999social,degroot1974reaching} and self-exciting event cascades \citep{Hawkes1971Spectra,Zhao2015SEISMIC,Rizoiu2018TutorialHawkes}---but these have rarely been brought to bear on the specific problem of AI governance stability. Friedkin--Johnsen-type models capture how beliefs propagate through networks under social influence and individual inertia, yet they do not account for the endogenous generation of trust-eroding events. Hawkes process models elegantly describe how events cluster and cascade, but treat triggering conditions as exogenous rather than coupling them to the cognitive states of the population.

To bridge this gap, we propose a baseline coupled dynamical framework for AI governance systems that jointly captures the co-evolution of public trust and AI-related controversy, integrating (i) a Friedkin--Johnsen trust propagation process in which individuals update their institutional trust under social influence, personal predisposition, and exposure to controversy, and (ii) a discrete-time Hawkes-inspired intensity process in which perceived controversy is endogenously modulated by the prevailing level of public trust. The framework is designed to demonstrate that the boundary between resilience and collapse is a structural property of the trust--event coupling, captured by an exact spectral criterion: only structural parameter changes can move the system across this boundary, but once the architecture lies in the unstable regime, a generic perturbation seeds a growing departure from equilibrium, while below the boundary the same shock dissipates asymptotically. The baseline model thus does not represent shock-induced regime switching: a controversy shock changes the \emph{state} of the system, whereas the spectral boundary is crossed only when architecture-defining \emph{parameters} change; what the model characterizes is recoverability under a fixed architecture, not irreversible tipping between governance regimes.

This paper makes the following contributions:
\begin{itemize}
    \item We propose a bidirectionally coupled dynamical model in which institutional trust and AI controversy events co-evolve through first-order endogenous feedback, using CASA-motivated social-update continuity as an explicit working hypothesis rather than an established equivalence.
    \item We derive the exact spectral stability criterion $\rho(J_{2n}) < 1$ that delineates recovery from growing departure under the baseline dynamics.
    \item We demonstrate through numerical experiments that event self-excitation and memory persistence systematically narrow the stable parameter regime, while echo-chambered topologies reshape distributional trust geometry under near-identical spectral boundaries.
    \item We identify four structural implications---fragile high-trust systems, stable low-trust regimes, the normative independence of fairness from dynamical stability, and bounded topology sensitivity in a memory-dominated regime---while stating the conditions under which each applies.
    \item We analyze a linear relaxation of the exogenous-event assumption in-paper (Section~\ref{sec:relaxA2}) and identify endogenous influence topology and fully endogenous event generation as follow-up extensions.
\end{itemize}

% ============================================================
% ===== Section 2: Related Work =====
% ============================================================
\section{Related Work}
\label{sec:related}

\subsection{Trust in AI and Institutional Governance}

Trust has emerged as a crucial mechanism for navigating intricacy in technology, organisations, and human interactions \citep{Lee2004Trust}. Following \citet{Mayer1995OrganizationalTrust}, we understand trust as the willingness of one entity (the trustor) to be vulnerable to another (the trustee), with the expectation that the trustee will execute an action critical to the trustor regardless of the trustor's capacity to oversee or control. We operationalize this notion as a time-evolving vector $T_t$ representing the trust levels of individual agents in a social network, aligning with the dynamic and socially constructed view of trust \citep{Mayer1995OrganizationalTrust,Lee2004Trust,Hoff2015,Rousseau1998Trust}.

The CASA literature shows that people often respond socially to interactive systems, but it does not establish that human--AI and interpersonal trust are identical \citep{Nass1994CASA,ReevesNass1996MediaEquation,NassMoon2000,LeeNass2010TrustCASA,Gambino2020CASA}. Evidence from avatars and automation documents both continuity and attenuation \citep{Riedl2014TrustingAvatars,Madhavan2007HumanAutomation}. Recent LLM experiments sharpen this mixed picture: anthropomorphic modality and wording can alter perceived accuracy and trust, while uncertainty language can reduce over-reliance, with effects depending on phrasing and context \citep{Cohn2024Anthropomorphism,Kim2024LLMUncertainty}. We therefore treat social influence as a modeling hypothesis whose parameters require LLM- and context-specific calibration; Section~\ref{sec:scope} develops the implications.

Existing AI governance scholarship has identified transparency, fairness, and accountability as widely recurring principles \citep{Jobin2019GlobalLandscape,Floridi2018AI4People}---principles that binding regulation has since begun to codify \citep{EUAIAct2024}---and has advanced institutional accountability mechanisms such as end-to-end algorithmic auditing \citep{Raji2020Accountability}, alongside critiques showing that formally fair algorithms can still fail in their surrounding sociotechnical context \citep{Selbst2019Fairness}. These approaches answer normative and institutional questions that a stability model cannot replace. Our narrower contribution is to supply an explicit recovery criterion for a coupled trust--controversy system, thereby complementing rather than formalizing fairness or legitimacy.

\subsection{Opinion Dynamics and Event Cascade Models}

Classical opinion dynamics begins with the French--DeGroot model \citep{degroot1974reaching} and was extended through exogenous influences \citep{abelson1964mathematical,taylor1968towards}, game-theoretic intrinsic opinions \citep{bindel2015how}, and the Friedkin--Johnsen (FJ) model with self-weighted ``stubbornness'' enabling persistent disagreement and polarization \citep{friedkin1990social,friedkin1999social,parsegov2016novel}. We adopt the FJ framework as the foundation for modeling trust propagation in this single-topic setting.

On the event-process side, homogeneous Poisson models \citep{Cox1966Statistical} are inadequate for the bursty, heavy-tailed patterns of real social behavior \citep{Barabasi2005Bursts}. Threshold and epidemic models \citep{Granovetter1978Threshold,Watts2002GlobalCascades} capture interpersonal influence but cannot represent the temporal clustering of spontaneous reactions. Hawkes processes \citep{Hawkes1971Spectra} address this by modeling self-exciting event rates, where each past event raises the likelihood of future events through an exponentially decaying kernel. They have been successfully applied to information cascades \citep{Zhao2015SEISMIC,Farajtabar2017Coevolve,Rizoiu2018TutorialHawkes}. We therefore adopt a discrete-time Hawkes-inspired formulation as the core of our event layer---specifically, its deterministic first-moment (intensity) form (Section~\ref{sec:model}).

\subsection{Coupled Trust--Event Dynamics}

Recent work has begun to address the dynamic interplay between trust and belief formation. \citet{ma2024social} propose a coupled decision-making model on directed graphs in which both trust relationships and individual opinions evolve simultaneously; \citet{martins2013trust} extend the Continuous Opinions and Discrete Actions framework with time-varying interpersonal trust weights. Beyond internal network dynamics, \citet{gallo2020social} investigate ``shock elections'' in a DeGroot framework with confirmation bias, where agents permanently sever links to sufficiently distant neighbors at the outset; the shock there is an emergent outcome of the learning process itself, and no ongoing event process is coupled to the belief state. The literature thus exhibits two complementary limitations: co-evolutionary models often neglect disruptive events or carry no attitude state at all (point-process models such as COEVOLVE, e.g., \citealp{Farajtabar2017Coevolve}, couple diffusion to network structure but not to trust), while models of information shocks do not endogenize the generation of controversy intensity from the trust state. Our framework integrates both perspectives via a hybrid FJ--Hawkes formulation in which trust and controversy intensity co-evolve through bidirectional first-order coupling.

\subsection{Scope and Social-Update Continuity}
\label{sec:scope}

The classical tools we employ were developed for interpersonal influence, whereas AI-mediated environments add algorithmic curation and a non-human trustee. The literature reviewed above motivates social influence as a plausible first-order update class, but it neither establishes the particular linear equation used here nor implies that human--AI and interpersonal trust update identically. We therefore treat the FJ mechanism as a context-dependent modeling hypothesis whose parameters require empirical estimation.

Two structural features of AI-mediated environments do exceed the classical scope: (i) \emph{endogenous influence topology}, where the influence matrix $W$ becomes a function of the trust state via algorithmic recommendation feedback; and (ii) \emph{partially endogenous event generation}, where a non-trivial fraction of trust-relevant events are themselves AI system outputs, rendering the Hawkes baseline rate $\boldsymbol{\mu}$ coupled to the trust state. We treat both as exogenous in the baseline model (Assumptions A1--A2 in Section~\ref{sec:model}); A2 is then partially relaxed, within a linear regime, in Section~\ref{sec:relaxA2}. This deliberate scoping enables closed-form equilibrium solutions and a formal spectral stability condition, isolating the role of first-order trust--event feedback as a first-order driver of governance fragility; Section~\ref{sec:future} returns to these structural extensions as the principal follow-up directions.

% ============================================================
% ===== Section 3: Model =====
% ============================================================
\section{Model}
\label{sec:model}

We propose a coupled dynamical system that models the co-evolution of public institutional trust in AI governance systems and the perceived intensity of AI-related controversy. The model captures two critical aspects: (1) the propagation of institutional trust---the public's confidence in the fairness, accountability, and legitimacy of AI-driven public decision-making systems---within a social network under both peer influence and exogenous shocks, and (2) the evolution of perceived controversy intensity---encompassing perceived algorithmic unfairness, resource allocation disputes, accountability failures, and related public backlash---shaped by prior trust states and its own self-exciting dynamics. These are perception states: the model does not measure the actual fairness, harm, or legitimacy of the underlying AI system.

\subsection{Trust Update Equation}

Let $T_t \in \mathbb{R}^{n}$ denote the vector of \emph{institutional trust} among $n$ agents at time $t$, where each component lies in the range $(0,2)$ and represents an individual's perceived confidence in AI systems deployed in public decision-making. It is not an algorithmic-fairness score or a welfare outcome. The trust update mechanism is defined as:

\begin{equation}
T_{t+1} = f(T_t, S_t) = AWT_t + (I - A)T_1 + BS_t
\end{equation}

\noindent where:
\begin{itemize}
    \item $A \in \mathbb{R}^{n \times n}$ is a diagonal matrix capturing each agent's susceptibility to social influence \citep{friedkin1990social,friedkin1999social}.
    \item $W \in \mathbb{R}^{n \times n}$ is the normalized influence matrix representing the topology of information diffusion---including social media platforms and algorithmic recommendation systems---whose structure determines whether trust signals are broadly shared or trapped within echo chambers.
    \item $T_1 \in \mathbb{R}^n$ is the initial (inherent) trust vector, capturing prior beliefs and baseline confidence in AI governance.
    \item $S_t \in \mathbb{R}^{n}$ is the perceived controversy intensity vector at time $t$, where each component encodes agent-specific perceptions of AI governance controversies. Because the event update equation couples $S_t$ with the heterogeneous trust vector $T_t$ through accumulated memory, $S_t$ evolves into a non-uniform vector even if initialized uniformly.
    \item $B \in \mathbb{R}^{n\times n}$ is a diagonal reactivity matrix; the product $BS_t$ captures two-stage heterogeneity: agents differ both in what they perceive ($S_t$) and in how strongly they react ($B$).
\end{itemize}

The first term $AWT_t$ captures network-internal diffusion of trust through social interactions. The second term $(I - A)T_1$ represents agents' tendency to revert to their innate beliefs, embodying cognitive rigidity. The additive term $BS_t$ introduces perturbations from controversy events.

\paragraph{Assumptions.} The framework as formulated relies on three scoping assumptions; A2 is partially relaxed in Section~\ref{sec:relaxA2}, and the remaining relaxations are discussed as future work (Section~\ref{sec:future}):

\noindent\textbf{A1 (Exogenous Influence Topology).} The matrix $W$ is time-invariant and does not depend on the trust state $T_t$.

\noindent\textbf{A2 (Exogenous Baseline Event Rate).} The baseline rate vector $\boldsymbol{\mu}$ (Eq.~2) is a constant, independent of $T_t$ and the event history.

\noindent\textbf{A3 (Constant Coupling Coefficients).} The coefficients $\alpha$, $\beta$, $\gamma$ and the matrices $A$, $B$ are time-invariant constants.

\subsection{Controversy Intensity Update Equation}

To capture the temporally clustered, reactive nature of AI governance controversies, we model controversy event intensity as a memory-based function of prior institutional trust and past events:

\begin{equation}
S_{t+1} = g(T_t, S_t) = \boldsymbol{\mu} + \sum_{i=0}^{t} \gamma^{t - i} \left( \alpha T_i + \beta S_i \right)
\end{equation}

\noindent where $\boldsymbol{\mu}$ is the baseline rate of spontaneous controversy occurrence, $\gamma \in (0,1)$ is the memory decay factor, and $\alpha, \beta$ are scalar parameters governing the influence of trust levels and prior events. The coupled dynamics between $T_t$ and $S_t$ are illustrated in Figure~\ref{fig:coupled_dynamics}.

\begin{figure}[t]
\centering
\resizebox{\columnwidth}{!}{%
\begin{tikzpicture}[
  node distance=1.0cm and 1.5cm,
  every node/.style={align=center, font=\footnotesize},
  box/.style={draw, rounded corners, minimum width=1.8cm, minimum height=0.7cm, font=\scriptsize},
  arrow/.style={-{Latex}, thick}
]
\node[box] (T1) {$T_1$};
\node[box, right=of T1] (Tt) {$T_t$};
\node[box, right=of Tt] (Ttp1) {$T_{t+1}$};
\node[box, below=of Tt] (St) {$S_t$};
\node[box, below=of Ttp1] (Stp1) {$S_{t+1}$};
\node[box, below=of St] (Ht) {$H_t$};
\node[box, right=of Ht] (Htp1) {$H_{t+1}$};

\draw[arrow, bend left=18] (T1) to node[midway, above, font=\scriptsize] {$(I{-}A)T_1$} (Ttp1);
\draw[arrow] (Tt) -- (Ttp1) node[midway, above, font=\scriptsize] {$AWT_t$};
\draw[arrow] (St) -- (Ttp1) node[pos=0.45, right, font=\scriptsize] {$BS_t$};
\draw[arrow, bend right=35] (Tt) to node[pos=0.55, left, font=\scriptsize] {$\alpha T_t$} (Htp1);
\draw[arrow] (St) -- (Htp1) node[pos=0.4, above, sloped, font=\scriptsize] {$\beta S_t$};
\draw[arrow] (Ht) -- (Htp1) node[midway, below, font=\scriptsize] {$\gamma H_t$};
\draw[arrow] (Htp1) -- (Stp1) node[midway, right, font=\scriptsize] {$\boldsymbol{\mu}{+}H_{t+1}$};
\end{tikzpicture}
}
\caption{Coupled dynamics between trust $T_t$ and perceived event intensity $S_t$. The memory state accumulates as $H_{t+1} = \gamma H_t + \alpha T_t + \beta S_t$ and the event intensity is read out as $S_{t+1} = \boldsymbol{\mu} + H_{t+1}$ (Eq.~2); events feed back into trust through $BS_t$ (Eq.~1).}
\label{fig:coupled_dynamics}
\end{figure}

This structure preserves the historical-influence accumulation of Hawkes-type processes while aligning with the discrete-time update of the Friedkin--Johnsen model. The bidirectional coupling between $T_t$ and $S_t$ establishes a closed feedback loop whose stability is a structural property of the joint architecture: depending on the spectral characteristics of $(A, W, B, \alpha, \beta, \gamma)$, the same perturbation can either dissipate over time or grow without bound---monotonically or with oscillatory growth, depending on the dominant eigenvalue---regardless of where the equilibrium itself lies.

\paragraph{Deterministic intensity convention.} Equation~(2) is a \emph{deterministic Hawkes-inspired intensity recursion}, not a stochastic point process: no random event counts are sampled, and past \emph{intensities}, rather than realized events, drive the excitation. The two are exactly linked. Consider the stochastic completion in which, given $S_t$, a realized event count $N_t$ satisfies $\mathbb{E}[N_t \mid S_t] = S_t$ (e.g., conditionally Poisson) and both the excitation and the trust update are driven by $N_t$; because all updates are linear, taking expectations reproduces Eqs.~(1)--(2) exactly. The deterministic system is thus the exact first-moment dynamics of that point process, and the stability results below govern its mean trajectory. Sampling realized events and analyzing fluctuations around the mean---which matter most near the stability boundary---are deferred to Section~\ref{sec:future}.

\paragraph{Modeling convention for $\alpha$.} We retain $\alpha \geq 0$, consistent with the Hawkes-process convention that event intensity is non-negative in past covariates, and use $\alpha$ as a reduced-form \emph{attention-and-salience} channel: the social amplification of risk framework motivates the possibility that visibility and information mediation amplify the societal consequences of a perturbation \citep{Kasperson1988SARF}, and in AI-specific settings violations of near-perfect automation expectations produce disproportionately strong trust reactions \citep{Dzindolet2003TrustReliance,Madhavan2007HumanAutomation}---without establishing a universal monotonic mapping from trust level to $\alpha$. The positive-$\alpha$ convention is therefore a modeling choice rather than an empirical commitment: the trust state enters as a salience-mediated amplifier, not a direct grievance generator, and this baseline conflates trust \emph{quality} with attention \emph{intensity}, deferring multi-dimensional disentanglement to future work (Section~\ref{sec:future}). The joint sign convention of the two coupling directions is coordinate-dependent: the reparameterization $T_t \mapsto c - T_t$ induces $(\alpha, B) \mapsto (-\alpha, -B)$ and yields a similar Jacobian with identical spectrum (Appendix~\ref{supp:reparam}); it does not license changing the sign of $\alpha$ alone while holding $B$ fixed.

Equations (1)--(2) constitute the hybrid FJ--Hawkes framework, capturing first-order trust--event endogeneity (trust enters via $\alpha$, events via $B$); second-order structural endogeneity in which $W$ or $\boldsymbol{\mu}$ depend on $T_t$ is deferred to Section~\ref{sec:future}. Table~\ref{tab:notations} summarizes the notation.

\begin{table}[t]
\centering
\caption{Summary of notation.}
\label{tab:notations}
\small
\begin{tabular}{ll}
\toprule
\textbf{Symbol} & \textbf{Description} \\
\midrule
$T_t$ & Institutional trust vector at time $t$, $\in \mathbb{R}^n$ \\
$S_t$ & Perceived controversy intensity vector, $\in \mathbb{R}^n$ \\
$H_t$ & Aggregated memory state, $\in \mathbb{R}^n$ \\
$A$ & Diagonal susceptibility matrix (FJ model) \\
$B$ & Diagonal reactivity matrix \\
$W$ & Influence/adjacency matrix encoding information \\
& diffusion topology \\
$T_1$ & Inherent trust vector \\
$\boldsymbol{\mu}$ & Baseline event rate vector \\
$\alpha,\beta$ & Trust-to-event and event self-excitation coeffs.\\
$\gamma$ & Memory decay factor, $\in (0,1)$ \\
$\hat{X}_t$ & Augmented state $[T_t^\top, H_t^\top]^\top \in \mathbb{R}^{2n}$ \\
\bottomrule
\end{tabular}
\end{table}

To further clarify the behavioral interpretation, exogenous influence on trust operates through two layers of heterogeneity. First, the perceived event intensity $S_t$ is agent-specific: because each agent carries a distinct trust history, the memory accumulation process produces individualized perceptions of event salience. Second, the diagonal matrix $B$ captures each agent's reactivity to their perceived event intensity, representing heterogeneity in emotional responsiveness, cognitive bias, or risk attitude.

% ============================================================
% ===== Section 4: Theoretical Analysis =====
% ============================================================
\section{Theoretical Analysis}
\label{sec:theory}

\subsection{Equilibrium Analysis}

To analyze the steady-state behavior of the trust--event system, we consider the equilibrium equations:
\begin{align}
T^* &= AWT^* + (I - A)T_1 + BS^* \\
S^* &= \boldsymbol{\mu} + \left(\alpha T^* + \beta S^*\right)\frac{1}{1 - \gamma}
\end{align}

\noindent Rearranging (3) yields $T^* = (I-AW)^{-1}(I-A)T_1 + (I-AW)^{-1}BS^*$, and rearranging (4) yields $(1-\beta/(1-\gamma))S^* = \boldsymbol{\mu} + (\alpha/(1-\gamma))T^*$. Defining
\[
X = (I-AW)^{-1}(I-A)T_1,\ \ Y = (I-AW)^{-1}B,
\]
\[
u = 1 - \frac{\beta}{1-\gamma},\ \ v = \frac{\alpha}{1-\gamma},
\]
the system reduces to $T^* = X + YS^*$ and $uS^* = \boldsymbol{\mu}+vT^*$, with closed-form solution
\begin{equation}
\begin{cases}
S^* = (uI - vY)^{-1}(\boldsymbol{\mu} + vX) \\
T^* = X + Y(uI - vY)^{-1}(\boldsymbol{\mu} + vX)
\end{cases}
\end{equation}

The invertibility of $(I - AW)$ and $(uI - vY)$ imposes explicit boundary conditions. First, $(I-AW)$ is non-singular whenever $\rho(AW) < 1$, which is guaranteed for row-stochastic $W$ and diagonal $A$ with $a_i \in (0,1)$. Second, the \emph{unique-equilibrium formula} loses invertibility where $\det(uI-vY)=0$; at such a point the affine equilibrium equations may have no solution or a non-unique solution, depending on compatibility of the constant term. This singular surface coincides exactly with $1 \in \sigma(J_{2n})$. Under the negative-reactivity convention $b_i < 0$ used throughout, a diagonal-dominance argument (Appendix~\ref{supp:equilibrium}) shows that the first such crossing lies strictly \emph{above} the memory-saturation line, at $\gamma + \beta = 1 + O(\alpha\, b_{\max})$; on the line $\gamma+\beta=1$ itself the scalar $u$ vanishes, but $(uI-vY)=-vY$ remains invertible and both $T^*$ and $S^*$ remain finite. Thus $\{\beta+\gamma\leq1,\ \rho(AW)<1\}$ is a guaranteed nonsingular region under $b_i<0$, not the maximal validity domain of Eq.~(5).

Eq.~(5) is contingent on A1--A2; under endogenous $W(T)$ or $\boldsymbol{\mu}(T)$, one possible existence route is Brouwer's theorem, provided the nonlinear update is continuous and maps a compact convex state set into itself (Section~\ref{sec:future}).

\subsection{Stability Analysis}
\label{sec:stability}

With the memory decay mechanism, the event intensity $S_{t+1}$ depends on a decaying accumulation of prior trust and event signals. This renders the system non-Markovian and prevents direct Jacobian-based analysis. We therefore apply state augmentation by introducing an auxiliary memory vector $H_t \in \mathbb{R}^n$ with the recursive update
\begin{equation}
H_{t+1} = \gamma H_t + \alpha T_t + \beta S_t,\quad S_{t+1} = \boldsymbol{\mu} + H_{t+1},\quad H_0 = \mathbf{0}.
\end{equation}

Since $S_t = \boldsymbol{\mu} + H_t$ for all $t \geq 1$, $S_t$ and $H_t$ are not independent. Substituting yields a non-redundant augmented state
\[
\hat{X}_t = [T_t^\top, H_t^\top]^\top \in \mathbb{R}^{2n},
\]
with component-wise update
\begin{align}
T_{t+1} &= AWT_t + BH_t + (I-A)T_1 + B\boldsymbol{\mu}, \\
H_{t+1} &= \alpha T_t + (\gamma + \beta)I_n H_t + \beta\boldsymbol{\mu}.
\end{align}
The system is affine with constant Jacobian
\begin{equation}
J_{2n} =
\begin{bmatrix}
AW & B \\
\alpha I_n & (\gamma + \beta) I_n
\end{bmatrix} \in \mathbb{R}^{2n \times 2n}.
\end{equation}
Because the dynamics are affine with a constant Jacobian, $\rho(J_{2n}) < 1$ is equivalent to \emph{global} asymptotic convergence to the equilibrium in the unconstrained state space; the substantive interpretation remains conditional on trajectories staying within the declared trust domain (Section~\ref{sec:experiments}).

Applying the Schur complement (Appendix~\ref{supp:eigenvalues}), the characteristic equation reduces to
\begin{equation}
\det\!\Big(AW - \tfrac{\alpha}{\gamma + \beta - \lambda} B - \lambda I_n\Big) = 0,\ \ \lambda \neq \gamma + \beta,
\end{equation}
a nonlinear eigenvalue problem valid for arbitrary network topologies and reactivity structures.

When $AW$ and $B$ are simultaneously diagonalizable (e.g., $B = bI_n$), the problem decouples into $n$ quadratics
\begin{equation}
(\lambda_k^{AW} - \lambda)(\gamma + \beta - \lambda) = \alpha b_k,\ \ k=1,\dots,n,
\end{equation}
with roots
\begin{equation}
\lambda_{k,\pm} = \frac{(\lambda_k^{AW}+\gamma+\beta) \pm \sqrt{(\lambda_k^{AW}-\gamma-\beta)^2 + 4\alpha b_k}}{2}.
\end{equation}
Instability may arise from any single mode---not necessarily the dominant one.

\paragraph{Real vs.\ complex eigenvalue regimes.} This analysis concerns the simultaneously diagonalizable case and, within it, modes with real $\lambda_k^{AW}$; the Perron eigenvalue of the non-negative matrix $AW$ is real and is therefore among the modes covered (for complex $\lambda_k^{AW}$ the two roots of Eq.~(11) are not conjugate and no single-root modulus formula exists; Appendix~\ref{supp:eigenvalues}). In the heterogeneous experiments of Section~\ref{sec:experiments}, stability is evaluated directly from the full Jacobian $J_{2n}$ rather than through this mode decomposition. The discriminant $(\lambda_k^{AW}-\gamma-\beta)^2 + 4\alpha b_k$ determines the qualitative character of the dynamics. Under the sign convention $b_k < 0$ (controversies erode trust, consistent with the asymmetry principle that negative events weigh more heavily on trust; \citealp{Slovic1993PerceivedRisk}), the discriminant becomes negative whenever $|4\alpha b_k| > (\lambda_k^{AW}-\gamma-\beta)^2$, in which regime the eigenvalues form a complex-conjugate pair and the corresponding mode \emph{oscillates} rather than monotonically growing or decaying. A direct calculation reduces the squared magnitude to the clean form
\begin{equation}
|\lambda_{k,\pm}|^2 = \lambda_k^{AW}(\gamma+\beta) + \alpha|b_k|,
\end{equation}
so that stability in the complex regime requires
\begin{equation}
\lambda_k^{AW}(\gamma+\beta) + \alpha|b_k| < 1.
\end{equation}
The trust--event coupling strength $\alpha|b_k|$ therefore contributes \emph{linearly to the squared modal magnitude} $|\lambda_{k,\pm}|^2$, alongside the network--memory product $\lambda_k^{AW}(\gamma+\beta)$. Writing the real and imaginary parts of $\lambda_{k,+}$ as $a_k$ and $b'_k$ (Appendix~\ref{supp:eigenvalues}), the angular frequency is
\begin{equation}
\omega_k = \operatorname{atan2}(b'_k,a_k),
\end{equation}
giving a quantitative prediction for the period $2\pi/\omega_k$ of the trust--event cycle---a structural prediction that could in principle be calibrated against scandal--rebound patterns in institutional-trust time series. The complex regime requires $|\lambda_k^{AW} - \gamma - \beta| < 2\sqrt{\alpha|b_k|}$; Section~\ref{sec:experiments} tests parameter scaling under which this condition is not met for the dominant mode, so the simulated divergence is monotone (real-eigenvalue), while the complex regime is a distinct prediction with oscillatory failure-mode signatures.

\paragraph{State shocks versus structural change.} A one-time state shock $d$ applied at time $\tau$ evolves as $\Delta \hat{X}_{\tau+k} = J_{2n}^{k}\, d$; when $\rho(J_{2n}) < 1$ it dissipates asymptotically, although near the boundary the modal recovery time $-1/\log\rho(J_{2n})$ is approximately $(1-\rho(J_{2n}))^{-1}$ and non-normal dynamics can produce transient amplification before decay. A finite state shock cannot change the spectral classification of the fixed affine system, because it does not change $J_{2n}$: crossing $\rho(J_{2n}) = 1$ requires a change in the architecture-defining parameters $(W, A, B, \alpha, \beta, \gamma)$. Instability is accordingly a diagnostic of failed recovery under a fixed architecture---unbounded departure that eventually leaves the model's interpretive domain---not a literal prediction of observed trust values.

Because the baseline dynamics are affine, ``stability'' throughout refers to global asymptotic convergence in the unconstrained state space under A1--A3, with substantive interpretation conditional on the trust domain $(0,2)$ (Section~\ref{sec:experiments}); under relaxed assumptions (Section~\ref{sec:future}) the dynamics become nonlinear, stability claims become local to each equilibrium, and $\rho(J_{2n})<1$ no longer guarantees global convergence.

\subsection{Relaxing A2: A Trust-Dependent Baseline Rate}
\label{sec:relaxA2}

Assumption A2 treats the baseline event rate $\boldsymbol{\mu}$ as constant. As a first-order relaxation, let the baseline rate respond linearly to the trust state,
\begin{equation}
\boldsymbol{\mu}(T_t) = \boldsymbol{\mu}_0 + E\,T_t, \qquad E = \mathrm{diag}(e_1, \dots, e_n),\quad e_i \geq 0,
\end{equation}
capturing AI systems whose output rate---and hence controversy surface---scales with deployment and use; the term $E T_t$ encodes a reduced-form association between current trust and the perceived controversy baseline (via exposure, reliance, and attention), not a claim that trust directly generates controversies. We adopt a \emph{contemporaneous readout} convention: the event intensity is the state readout $S_t = \boldsymbol{\mu}_0 + E T_t + H_t$, so that after updating $(T_{t+1}, H_{t+1})$ the next intensity is read out as $S_{t+1} = \boldsymbol{\mu}_0 + E T_{t+1} + H_{t+1}$---the current trust state shapes the current controversy surface rather than acting with a one-period lag. The system remains linear--affine: substituting this readout into Eqs.~(1) and (6) yields an augmented system on $(T_t, H_t)$ with Jacobian
\begin{equation}
J_{2n}^{E} =
\begin{bmatrix}
AW + BE & B \\
\alpha I_n + \beta E & (\gamma + \beta) I_n
\end{bmatrix},
\end{equation}
so the entire apparatus of Section~\ref{sec:stability} carries over with effective network block $AW + BE$ and effective coupling column $\alpha I_n + \beta E$; the Schur reduction holds verbatim because all new blocks are diagonal, and when $AW$, $B$, and $E$ are simultaneously diagonalizable the mode quadratic becomes $(\lambda_k^{AW} + b_k e_k - \lambda)(\gamma + \beta - \lambda) = (\alpha + \beta e_k)\, b_k$. Two consequences follow. First, in the complex regime (real-$\lambda_k^{AW}$ modes) the squared modulus shifts by exactly $-\gamma\, e_k |b_k|$ relative to the baseline: under negative reactivity, a trust-responsive baseline rate is \emph{stabilizing}---the erosion term $BE \leq 0$ dampens the effective network gain faster than the added coupling $\beta e_k b_k$ feeds back. Numerically, $\rho(J_{2n}^{E})$ decreases monotonically in the response strength for the baseline ensemble (uniform $e_i = e$: from $0.8999$ at $e=0$ to $0.8990$ at $e=0.1$ and $0.8981$ at $e=0.2$). Second, Theorem~S1 extends to $J_{2n}^{E}$ with $M$ replaced by the bound $\max_i (a_i + |b_i| e_i)$ and $q$ by $b_{\max}(\alpha + \beta e_{\max})$; the extension rescales the two off-diagonal coupling blocks jointly and is proven in the extension remark of Appendix~\ref{supp:topology}. The qualitatively new phenomena expected from event endogeneity---multi-type cascades and nonlinear response---require $\boldsymbol{\mu}(T)$ beyond the linear regime and are deferred to Section~\ref{sec:future}.

% ============================================================
% ===== Section 5: Numerical Experiments =====
% ============================================================
\section{Numerical Experiments}
\label{sec:experiments}

\subsection{Simulation Setup}

Setup: $n=5$ agents, $T=50$ steps; $T_i(0), T_{1,i} \sim \mathcal{U}(0,2)$, $S_0 = 0.1\,\mathbf{1}$; base parameters $\mu=0.1, \alpha=0.005, \beta=0.4, \gamma=0.5$; matrices $a_i \sim \mathcal{U}(0.4,0.9)$, $W_{ij}\sim\mathcal{U}(0,1)$ for $i \neq j$ with $W_{ii}=0$ (row-normalized), and $B_i \sim \mathcal{U}(-0.05,-0.01)$ (strictly negative reactivity, consistent with the controversy-erodes-trust narrative throughout).

\begin{figure*}[t]
    \centering
    \includegraphics[width=0.85\textwidth]{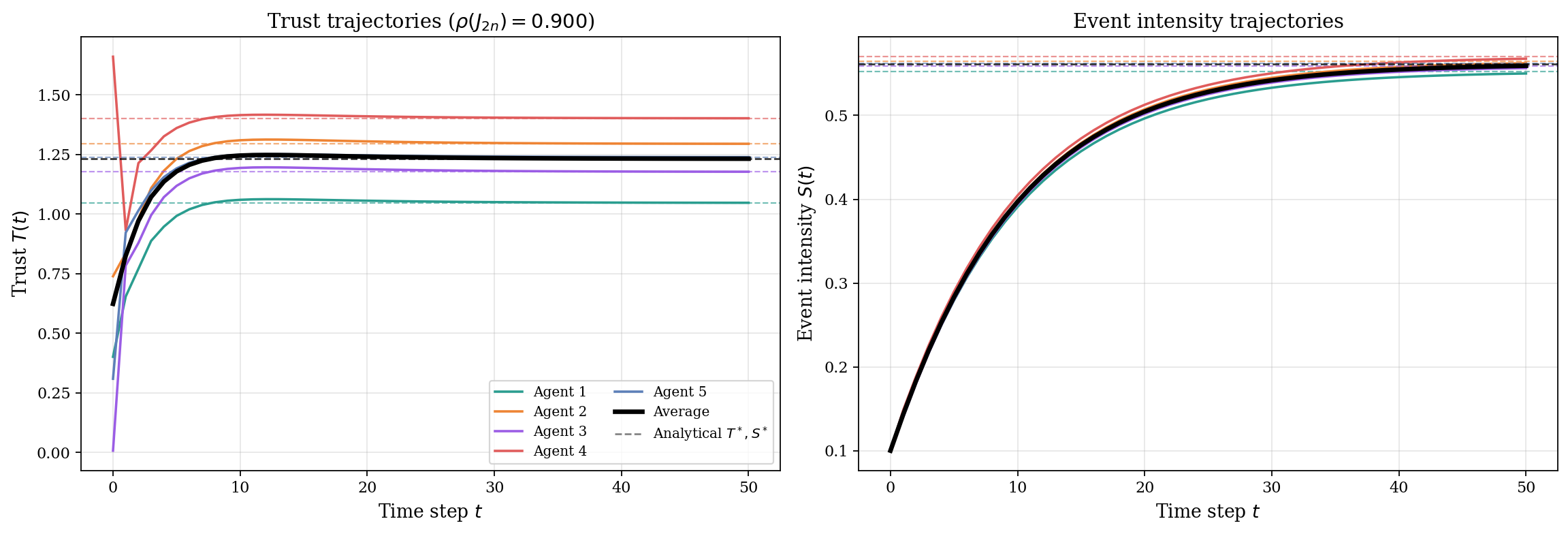}
    \caption{Simulated trust and event trajectories ($T=50$). Dashed horizontal lines mark the analytical fixed point $(T^*, S^*)$ from Eq.~(5); the simulated dynamics converge onto it.}
    \label{fig:sim50}
\end{figure*}

\subsection{Observations}

Trust trajectories converge to the analytical fixed point after a short transient overshoot (Fig.~\ref{fig:sim50}), consistent with the real-eigenvalue regime (Section~\ref{sec:stability}) under the parameter scaling chosen here.

\subsection{Sensitivity Analysis}

\paragraph{Impact of $\bm{\alpha}$.} Varying $\alpha \in \{0, 0.005, \dots, 0.5\}$ with $\mu=0.1, \beta=0.4, \gamma=0.5$ fixed, we observe a monotonic relationship: equilibrium event intensity rises and average equilibrium trust falls as the coupling strengthens. The system remains stable across the scanned range: direct computation of the full Jacobian over the sweep finds $\rho(J_{2n}) \leq \gamma+\beta = 0.9$ throughout, with the maximum attained at $\alpha = 0$---under negative reactivity, the real-regime effect of increasing $\alpha$ on the dominant eigenvalue is mildly stabilizing. Sufficiently large $\alpha$---well beyond the range examined here---does destabilize the system through the complex-regime channel of Section~\ref{sec:stability}.

\begin{figure}[t]
    \centering
    \includegraphics[width=0.95\columnwidth]{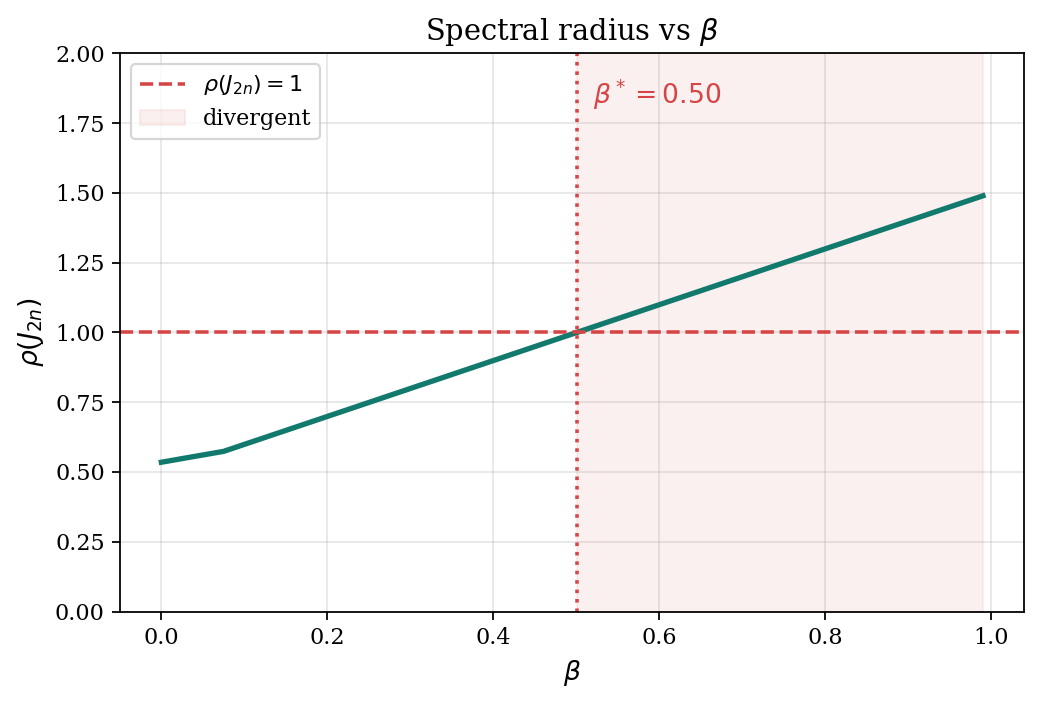}
    \caption{Sensitivity of $\beta$ ($\gamma=0.5,\ \alpha=0.05$). The dashed line marks $\rho(J_{2n})=1$, crossed at $\beta^*\approx 0.50$, and the shaded region is divergent. Each point is a separate fixed-parameter system: the sweep traces a boundary in parameter space, not a shock-induced transition within one trajectory.}
    \label{fig:betaSim}
\end{figure}

\paragraph{Impact of $\bm{\beta}$.} Varying $\beta \in [0,1]$ with $\mu=0.1, \alpha=0.05, \gamma=0.5$ and $B_i \sim \mathcal{U}(-0.05,-0.01)$ (controversies erode trust, consistent with the narrative in Section~\ref{sec:model}; the sensitivity sweeps raise $\alpha$ from its baseline $0.005$ to $0.05$ to make coupling effects visible, which shifts the boundary location by less than $10^{-3}$), the system exhibits a stability threshold at $\beta^* \approx 0.50$ where $\rho(J_{2n})=1$ (Fig.~\ref{fig:betaSim}). For $\beta<\beta^*$ the system converges; beyond $\beta^*$ it diverges, illustrating the structural collapse mechanism: an architectural change in event self-excitation, not a change in trust level, drives the system across the stability boundary. The scaling here places the dominant mode in the real-eigenvalue regime, so divergence past $\beta^*$ is monotone (not the complex regime's oscillatory envelope, Section~\ref{sec:stability}).

\paragraph{Impact of $\bm{\gamma}$.} Varying $\gamma \in [0,1]$ with $\mu=0.1,\alpha=0.05,\beta=0.3$, we again observe a stability boundary at $\gamma^* \approx 0.70$ (with $\beta=0.3$; the topology experiment of Section~5.4 uses $\beta=0.35$ and correspondingly finds $\gamma^* \approx 0.65$---both instances of the same leading-order relation $\gamma+\beta \approx 1$). Holding the other sampled quantities fixed, longer memory horizons raise the equilibrium event intensity and eventually move this parameterized system across the structural stability boundary.

\paragraph{Domain validity near the boundary.} Immediately inside the spectral boundary (numerically, $\beta \in [0.497,\beta^*)$ with $\beta^*\approx0.501$ at $\gamma=0.5$) the system still converges, but the equilibrium trust components exit the interpretive range $(0,2)$ declared in Section~\ref{sec:model}, and convergence times grow approximately as $(1-\rho(J_{2n}))^{-1}$. Stable-side statements should therefore be read at a finite margin from the boundary---the standard caveat for linear dynamics used outside a saturating domain.

\subsection{Effect of Network Topology}

The preceding analyses fix $W$ and vary scalar parameters. In practice, the structure of social influence itself plays a critical role. We compare three topologies with $n=10$ agents under identical coupling parameters ($\alpha=0.05,\beta=0.35,\gamma=0.5$) and strictly negative reactivity $B_i \sim \mathcal{U}(-0.05,-0.01)$ consistent throughout:

\begin{itemize}
    \item \textbf{Random network.} Off-diagonal entries drawn from $\mathcal{U}(0,1)$ and row-normalized; baseline with no preferential structure.
    \item \textbf{Echo chamber.} Two equally sized groups; intra-group weights $\sim\mathcal{U}(0.5,1.0)$, inter-group weights $\sim\mathcal{U}(0.01,0.05)$, then row-normalized.
    \item \textbf{Star / KOL.} A single hub listens to all peripherals; each peripheral assigns $\sim 80\%$ of its influence weight to the hub.
\end{itemize}

\begin{figure*}[t]
    \centering
    \includegraphics[width=0.95\textwidth]{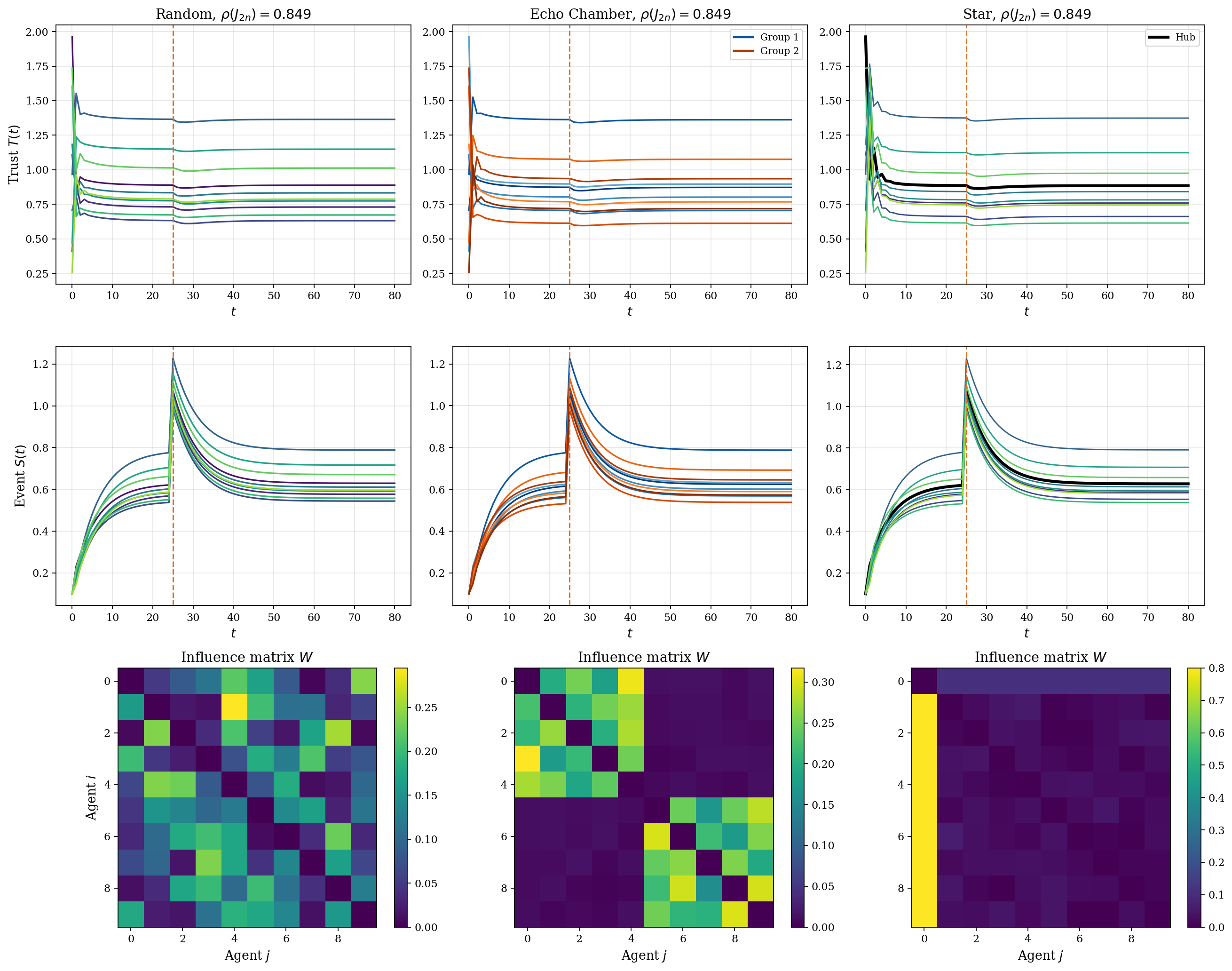}
    \caption{Effect of network topology on trust--event dynamics ($n=10$), with identical draws of $A$, $B$, $T_0$, $T_1$ across columns. Columns: Random, Echo Chamber (trajectories colored by group), Star/KOL (hub in black). Top: trust trajectories; middle: event intensities; bottom: $W$ heatmap. A one-step pulse of magnitude $\Delta H=0.45$ is applied to event memory at $t=25$ (vertical dashed line); all three systems return toward equilibrium, consistent with $\rho(J_{2n})=0.849<1$. The pulse tests recovery under a fixed Jacobian and does not change the spectral regime.}
    \label{fig:topoCompare}
\end{figure*}

In the random network, trust trajectories converge smoothly and group membership is immaterial (mean inter-group equilibrium gap $\approx 0.01$). In the echo chamber, the same agents organize into group-aligned bands whose means remain separated at equilibrium (mean inter-group gap $\approx 0.11$, an order of magnitude larger), because weak inter-group links prevent corrective cross-flow. In the star network, the hub's inherent trust exerts an outsized pull on the peripheral equilibria (Fig.~\ref{fig:topoCompare}).

\begin{figure*}[t]
    \centering
    \includegraphics[width=0.9\textwidth]{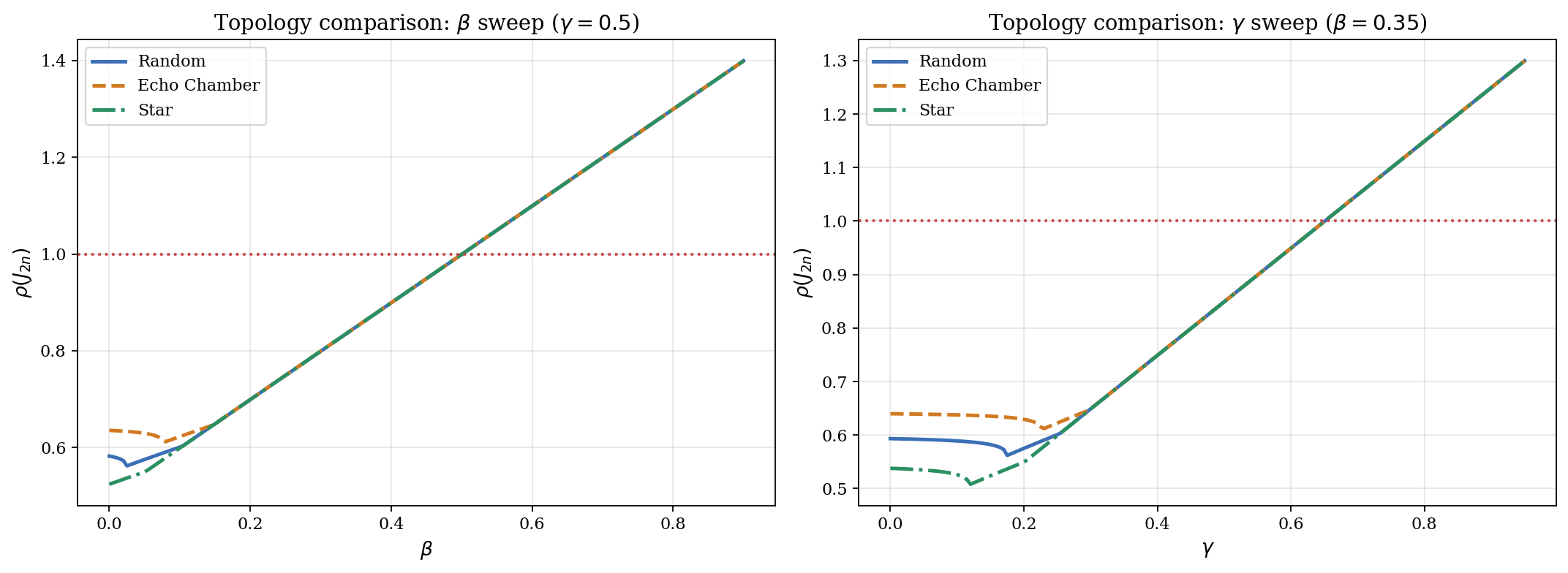}
    \caption{Topology sensitivity across the $\beta$ sweep ($\gamma=0.5$; critical $\beta^*\!\approx\!0.50$) and the $\gamma$ sweep ($\beta=0.35$; critical $\gamma^*\!\approx\!0.65$). The three curves separate at lower memory strength, confirming that fixed $W$ enters the spectrum, but nearly overlap as the common memory term dominates near the reported boundary. The latter is the empirical signature of the conditional topology--stability bound in Implication~4 and Theorem~S1, not universal topology invariance.}
    \label{fig:topoStability}
\end{figure*}

Fig.~\ref{fig:topoStability} reports $\rho(J_{2n})$ vs $\beta$ and $\gamma$ for the three topologies. Under the chosen parameters, all three yield nearly identical spectral boundaries ($\beta^* \approx 0.50$, $\gamma^* \approx 0.65$). This is the empirical manifestation of the \emph{approximate topology--stability decoupling} we name as Implication 4 of Section~\ref{sec:counterintuitive}: topology shapes \emph{where the equilibrium lies}, while the common memory term dominates the spectral radius in this parameter regime. Fig.~\ref{fig:topoCompare} makes the equilibrium-side effect visible---broad mixing under random influence, persistent cluster separation in the echo chamber, and hub-mediated centralization under the star---all under identical coupling parameters and baseline trust. Appendix~\ref{supp:topology} complements this with a theorem-aligned ensemble ($a_i \leq 0.70$) in which the sufficient condition of Theorem~S1 holds with margin. This is not a universal invariance claim: because $W$ appears in $J_{2n}$, a fixed exogenous topology can materially affect the spectral radius, and can change the stability classification, outside the theorem's memory-dominated regime. Endogenous $W(T_t)$ adds a further state-dependent and nonlinear topology channel (Section~\ref{sec:future}); it is not required for topology to matter.

% ============================================================
% ===== Section 6: Discussion =====
% ============================================================
\section{Discussion}
\label{sec:discussion}

\subsection{Four Structural Implications of the Baseline Model}
\label{sec:counterintuitive}

The repositioning of stability as a structural rather than a level property yields four implications of the baseline model that depart from common intuition in AI governance discourse. We state them as model implications---not empirical findings---because each rests on a conceptual confusion that the formal analysis cleanly resolves.

\paragraph{Implication 1: High-trust AI systems can be structurally fragile.} The dominant policy intuition---``if the public trusts the AI, governance is safe''---conflates trust \emph{level} with system \emph{robustness}. The equilibrium $T^*$ depends on the affine inputs, including $T_1$, whereas the spectral radius $\rho(J_{2n})$ controlling dissipation is independent of $T_1$ and is set by $(W, A, B, \alpha, \beta, \gamma)$. A high-trust equilibrium can therefore lie close to or beyond the stability boundary: a small margin $1-\rho(J_{2n})$ implies slow recovery and potentially large transient responses, while sustained growth requires $\rho(J_{2n})\geq1$ already or a structural parameter change that crosses the boundary. This is a structural possibility, not a calibrated empirical claim.

\paragraph{Implication 2: Low-trust environments can be structurally stable.} The complementary misconception holds that low public trust is itself a sign of dynamical instability. Our framework distinguishes between a \emph{low equilibrium} ($T^*$ small) and an \emph{unstable equilibrium} ($\rho(J_{2n}) \geq 1$): these are independent properties of the coupled system. When event self-excitation is weak and memory decays quickly---so that the resulting spectral radius lies below one---a low-trust regime can remain dynamically stable: public skepticism without growing departure from equilibrium. Such stability describes recovery to a low-trust equilibrium; it does not imply that the governance arrangement is legitimate, desirable, or successful.

\paragraph{Implication 3: Dynamical stability is not an algorithmic-fairness result.} Fairness is not a state or parameter in the baseline model, so the analysis cannot rank deployments by fairness or prove how a fairness intervention changes stability. A narrower conditional statement is available. Any intervention represented only by an affine input such as $T_1$ or $\boldsymbol{\mu}$ changes the equilibrium but not $J_{2n}$; an intervention that changes $A$, $B$, $W$, $\alpha$, $\beta$, or $\gamma$ can change the spectral radius and must be analyzed through its actual parameter pathway. Assigning a real fairness intervention to either class requires empirical and institutional evidence that the present model does not supply. Thus fairness and recoverability are complementary evaluative dimensions, not substitutes: sociotechnical fairness analysis asks how benefits, burdens, and harms are distributed \citep{Selbst2019Fairness}, while end-to-end auditing examines the broader organizational process that produces those outcomes \citep{Raji2020Accountability}; the present criterion asks only whether deviations decay under a specified dynamics.

\paragraph{Implication 4: Topology sensitivity is regime-dependent.} A substantial literature documents homophilous segregation in online information ecosystems and its risks for democratic discourse \citep{Sunstein2017Republic,Cinelli2021EchoChamber}. Our analysis separates topology's effect on the equilibrium from its effect on asymptotic stability. The matrix $W$ always enters both $(I-AW)^{-1}$ and $J_{2n}$; therefore even a fixed exogenous topology can affect the spectral radius. Theorem~S1 establishes the more specific result that, when the common memory term dominates by the stated margin, this spectral effect is uniformly bounded: $\rho(J_{2n})$ remains within $\delta_-\leq\sqrt{\alpha b_{\max}}$ of $\gamma+\beta$ for every row-stochastic $W$. The experiments illustrate this memory-dominated regime: random, echo-chamber, and star topologies produce visibly different equilibrium distributions but nearly identical spectral boundaries. Outside the theorem's premise, fixed $W$ can matter substantially and can place different topologies on opposite sides of $\rho=1$; making $W$ endogenous adds another nonlinear channel but is not the only way topology affects stability.

Together, these four implications distinguish trust level, recoverability, normative quality, and topology sensitivity rather than collapsing them into a single notion of ``trustworthy governance.'' Section~\ref{sec:structural-interventions} states how the stability analysis can complement, but not replace, fairness and accountability assessment.

\subsection{Recoverability Is Neither Fairness nor Legitimacy}
\label{sec:structural-interventions}

Recovery of public trust does not imply that an AI system deserves that trust. A rapid return to equilibrium may reflect repair, but it may instead mean that dissent following bias, privacy violations, or procedural unfairness dissipated without redress; a stable low-controversy state may likewise coexist with suppressed grievances. The baseline contains no fairness metric, protected-group outcome, or harm state, and its stability diagnostics should not be read as governance endorsements.

Assessment must therefore remain two-part. Fairness and accountability analysis should evaluate distributions of benefit and harm, procedural rights, and affected groups' experience \citep{Selbst2019Fairness,Raji2020Accountability}; stability analysis separately asks whether deviations decay under the specified architecture. A change confined to an affine input moves the equilibrium without changing asymptotic stability, whereas a change to a Jacobian parameter can affect both, but mapping any real intervention to either class requires empirical evidence. Future extensions should explicitly represent harm exposure, voice, contestation, and institutional response rather than relabeling trust parameters as fairness.

\subsection{Limitations}

\paragraph{Primary scoping limitation: social-update continuity and structural features of AI mediation.} The framework uses social-update continuity as a working hypothesis for trust micro-mechanics (Section~\ref{sec:scope}) while treating two structural features of AI-mediated environments as exogenous. LLM-era experiments show that modality, anthropomorphic framing, and uncertainty language can alter trust and reliance in context-specific ways \citep{Cohn2024Anthropomorphism,Kim2024LLMUncertainty}; the linear FJ update does not identify those mechanisms separately. In addition, $W$ is fixed under A1 even though algorithmic curation may make it trust-dependent, and the deterministic Hawkes-inspired event layer has an exogenous baseline rate under A2. Section~\ref{sec:relaxA2} takes a linear first step toward a trust-dependent baseline; nonlinear, multi-type, and stochastic cases remain open.

\paragraph{Secondary limitations.} The dynamics are linear, missing saturation, tipping, and hysteresis effects that real-world trust likely exhibits. The model is deterministic, omitting stochastic shocks. Parameters have not yet been empirically calibrated; calibration against longitudinal survey or social-media data is a prerequisite for predictive use.

\subsection{Future Directions}
\label{sec:future}

The baseline framework opens follow-up work in structural extensions, affected-group outcomes, and empirical grounding.

\paragraph{Structural Extensions.}
(i) \emph{From exogenous to endogenous influence topology.} Relaxing Assumption A1 amounts to specifying a topology-update law $W_{ij}(t+1) = \mathcal{M}(W_{ij}(t), T_i(t), T_j(t))$, which can take similarity-based forms ($W_{ij} \propto \exp(-\kappa|T_i - T_j|)$, modeling algorithmic homophily), engagement-weighted forms, or hybrids. Because fixed $W$ already appears in $J_{2n}$, endogeneity is not required for topology to influence stability; it adds a distinct feedback channel. At an equilibrium $T^*$, the linearization of $A W(T)T$ in a perturbation direction $\Delta T$ contains $A W(T^*)\Delta T + A\,DW(T^*)[\Delta T]\,T^*$. The second term is absent from the baseline and makes stability equilibrium-dependent. Sufficiently strong homophily may then produce multiple equilibria, bifurcation, or hysteresis, but these phenomena must be derived from a specified update law rather than presumed. (ii) \emph{From linear to fully endogenous events.} Section~\ref{sec:relaxA2} relaxes A2 in the linear regime; nonlinear response functions $\boldsymbol{\mu}(T_t)$ and multi-type Hawkes formulations in which controversy events are themselves typed AI-system outputs remain open---as does the fully stochastic point-process version, of which Eqs.~(1)--(2) are the exact mean dynamics (Section~\ref{sec:model}), together with the fluctuation-driven risk it implies near the stability boundary. (iii) \emph{From scalar to multi-dimensional trust.} Decomposing trust into technical capability, value alignment, and institutional embedding would enable richer analyses with partially independent dynamics per dimension.

\paragraph{Fairness and affected-group outcomes.}
A fairness-aware extension should introduce explicit outcome and harm variables, affected-group membership, and institutional response rather than relabeling trust parameters as fairness. Such a model could test how unequal exposure, voice, contestation capacity, and redress interact with perceived controversy and recovery. Only after those pathways are specified and calibrated would it be meaningful to ask whether a particular fairness intervention changes $T_1$, $B$, $\boldsymbol{\mu}$, or another component of the dynamics.

\paragraph{Empirical Calibration and Synthetic Stress Testing.}
Beyond structural and institutional extensions, the framework requires empirical grounding through longitudinal survey, platform, or administrative data. LLM-based agent simulations can provide complementary synthetic stress tests in which structural parameters are varied under controlled prompts, but they do not constitute empirical validation of public behavior. Their appropriate role is to probe model sensitivity and generate hypotheses that are then tested against observations of affected populations.

% ============================================================
%  Appendices (formerly the supplementary document S1--S5)
% ============================================================
\appendix

% ============================================================
\section{Equilibrium Derivation}
\label{supp:equilibrium}

We derive the closed-form equilibrium solution to the trust--event system defined by the vector equations:
\begin{align}
T^* &= AWT^* + (I - A)T_1 + BS^* \tag{S1.1}\\
S^* &= \boldsymbol{\mu} + \left(\alpha T^* + \beta S^*\right)\frac{1}{1 - \gamma} \tag{S1.2}
\end{align}

\paragraph{Step 1: Rearrange (S1.1).}
Isolating $T^*$ on the left-hand side:
\begin{align*}
(I - AW)T^* &= (I - A)T_1 + BS^* \\
\Rightarrow T^* &= (I - AW)^{-1}(I - A)T_1 \\
                & \quad + (I - AW)^{-1}B\,S^*.
\end{align*}

\paragraph{Step 2: Rearrange (S1.2).}
Solving for $S^*$ explicitly:
\begin{align*}
S^* &= \boldsymbol{\mu} + \frac{\alpha}{1 - \gamma}T^* + \frac{\beta}{1 - \gamma}S^* \\
\Rightarrow \left(1 - \tfrac{\beta}{1 - \gamma}\right) S^* &= \boldsymbol{\mu} + \tfrac{\alpha}{1 - \gamma} T^*.
\end{align*}

\paragraph{Step 3: Substitution.}
Define
\[
X = (I - AW)^{-1}(I - A)T_1,\quad Y = (I - AW)^{-1}B,
\]
\[
u = 1 - \tfrac{\beta}{1 - \gamma},\quad v = \tfrac{\alpha}{1 - \gamma}.
\]
The system reduces to
\[
T^* = X + Y S^*, \quad u S^* = \boldsymbol{\mu} + v T^*.
\]

\paragraph{Step 4: Solve.}
Substitute $T^* = X + Y S^*$ into the second equation:
\begin{align*}
u S^* &= \boldsymbol{\mu} + v(X + Y S^*) \\
\Rightarrow (uI - vY) S^* &= \boldsymbol{\mu} + vX.
\end{align*}
Assuming $(uI - vY)$ is invertible:
\[
S^* = (uI - vY)^{-1}(\boldsymbol{\mu} + vX),
\]
and substituting back:
\[
T^* = X + Y(uI - vY)^{-1}(\boldsymbol{\mu} + vX).
\]

\paragraph{Boundary conditions.}
The decomposition requires:
\begin{enumerate}
\item $(I - AW)$ non-singular, guaranteed when $\rho(AW) < 1$. For row-stochastic $W$ and diagonal $A$ with $a_i \in (0,1)$, we have $\rho(AW) \leq \max(a_i) < 1$. \checkmark
\item $(uI - vY)$ non-singular. Since $v = \alpha/(1-\gamma) > 0$, singularity requires $u/v \in \sigma(Y)$ with $Y = (I-AW)^{-1}B$. Under the negative-reactivity convention ($B$ diagonal, $b_i < 0$), $Y$ has no eigenvalue on the positive real axis: if $Yx = sx$ with $s > 0$, then $(sI - B - sAW)x = 0$; writing $-B = |B|$, the matrix $sI + |B| - sAW$ is strictly row-diagonally dominant---row $i$ carries diagonal weight at least $s + |b_i| - s a_i W_{ii}$ against off-diagonal mass $s a_i (1 - W_{ii})$, a margin of $s(1-a_i) + |b_i| > 0$---hence non-singular, a contradiction. Consequently $\det(uI - vY) \neq 0$ for every $u \geq 0$: for $u > 0$ the candidate eigenvalue $u/v$ is positive and excluded above, and at $u = 0$ the matrix $-vY$ is invertible because $B$ is. Genuine blow-up occurs only for $u < 0$ (i.e., $\beta + \gamma > 1$), where $u/v$ can meet a negative real eigenvalue of $Y$; moreover $\det(uI - vY) = 0$ is equivalent to $\det(I_{2n} - J_{2n}) = 0$, so the blow-up surface is exactly the surface on which the eigenvalue $+1$ enters $\sigma(J_{2n})$, located at $\gamma + \beta = 1 + O(\alpha b_{\max})$.
\end{enumerate}

The valid parameter region for the closed form is therefore $\{(\beta, \gamma): \beta + \gamma \leq 1\} \cap \{\rho(AW) < 1\}$ under $b_i < 0$, as stated in Section~4.1.

% ============================================================
\section{Jacobian Matrix Derivation}
\label{supp:jacobian}

\subsection{Non-redundant Jacobian}

Section~\ref{sec:stability} introduces the memory state $H_{t+1}=\gamma H_t+\alpha T_t+\beta S_t$. Since $S_t=\boldsymbol{\mu}+H_t$ for $t\geq1$, the non-redundant state $\hat{X}_t=[T_t^\top,H_t^\top]^\top$ obeys
\begin{align*}
T_{t+1} &= AWT_t + BH_t + (I-A)T_1 + B\boldsymbol{\mu}, \tag{S2.1}\\
H_{t+1} &= \alpha I_n T_t + (\gamma+\beta) I_n H_t + \beta\boldsymbol{\mu}. \tag{S2.2}
\end{align*}
Differentiating this affine map gives
\begin{equation}
J_{2n} =
\begin{bmatrix}
AW & B \\
\alpha I_n & (\gamma + \beta) I_n
\end{bmatrix} \in \mathbb{R}^{2n \times 2n}. \tag{S2.3}
\end{equation}
Under A1--A3 this Jacobian is constant, so its spectrum determines global asymptotic stability in the unconstrained state space.

\subsection{Equivalence with the Redundant Representation}

If one instead retains the redundant state $\tilde{X}_t = [T_t^\top, S_t^\top, H_t^\top]^\top \in \mathbb{R}^{3n}$ and uses $T_{t+1} = AWT_t + BS_t + \text{const}$ (with $S_t$ entering directly rather than $H_t$), the Jacobian takes the form:
\begin{equation}
J_{3n} =
\begin{bmatrix}
AW & B & \mathbf{0}_{n \times n} \\
\alpha I_n & \beta I_n & \gamma I_n \\
\alpha I_n & \beta I_n & \gamma I_n
\end{bmatrix}. \tag{S2.4}
\end{equation}

The second and third block rows are identical because both $S_{t+1}$ and $H_{t+1}$ satisfy the same recurrence in the variables $(T_t, S_t, H_t)$. The precise spectral relationship follows from a determinant identity.

\paragraph{Determinant identity.} Change coordinates by $(T, S, H) \mapsto (T, S-H, H)$, realized by the invertible block matrix $Q$ whose middle block row is $[\,\mathbf{0}\ \ I_n\ \ {-I_n}\,]$ and which is the identity elsewhere. Conjugation gives
\[
Q J_{3n} Q^{-1} =
\begin{bmatrix}
AW & B & B \\
\mathbf{0} & \mathbf{0} & \mathbf{0} \\
\alpha I_n & \beta I_n & (\beta+\gamma) I_n
\end{bmatrix},
\]
whose middle block row vanishes identically---the difference $S_{t+1} - H_{t+1} = \boldsymbol{\mu}$ is a constant, so its linearization is zero. Laplace expansion of $\det(Q J_{3n} Q^{-1} - \lambda I_{3n})$ along the middle block row (whose only nonzero block is $-\lambda I_n$) yields
\begin{equation}
\det(J_{3n} - \lambda I_{3n}) = (-\lambda)^n \det(J_{2n} - \lambda I_{2n}). \tag{S2.5}
\end{equation}

\paragraph{Conclusion.} The redundant representation carries exactly $n$ additional zero eigenvalues (with algebraic multiplicity), and its remaining spectrum coincides with that of $J_{2n}$, multiplicities included---uniformly across all degenerate cases (singular $B$, repeated rows of $W$).

% ============================================================
\section{Characteristic Polynomial and Eigenvalues}
\label{supp:eigenvalues}

\subsection{General Spectral Equation}

To determine the eigenvalues of $J_{2n}$, we solve
\begin{align*}
&\det(J_{2n} - \lambda I_{2n}) \\
&\quad = \det\!\begin{pmatrix} AW - \lambda I_n & B \\ \alpha I_n & (\gamma+\beta-\lambda) I_n \end{pmatrix} = 0.
\end{align*}
Since the lower-right block $(\gamma+\beta-\lambda)I_n$ is a scalar multiple of the identity, we apply the Schur complement formula. For $\lambda \neq \gamma + \beta$, the lower-right block is invertible with inverse $(\gamma+\beta-\lambda)^{-1} I_n$, and:
\begin{align}
&\det(J_{2n} - \lambda I_{2n}) \notag \\
&\quad = (\gamma + \beta - \lambda)^n \det\!\Big[AW - \lambda I_n - \tfrac{\alpha}{\gamma+\beta-\lambda} B\Big] = 0. \tag{S3.1}
\end{align}
The factorization is valid for $\lambda \neq \gamma+\beta$; the characteristic polynomial itself extends continuously as $p(\lambda) = \det[(\gamma+\beta-\lambda)(AW - \lambda I_n) - \alpha B]$, with $p(\gamma+\beta) = (-\alpha)^n \det B$, so $\lambda = \gamma+\beta$ is an eigenvalue of $J_{2n}$ if and only if $\alpha = 0$ or $B$ is singular---neither of which occurs in the parameter domain considered here. The second factor defines the core nonlinear eigenvalue problem:
\begin{equation}
\det\!\left( AW - \tfrac{\alpha}{\gamma+\beta-\lambda} B - \lambda I_n \right) = 0. \tag{S3.2}
\end{equation}
This is valid for arbitrary network topologies $W$ and reactivity structures $B$, with no diagonalizability assumption on $AW$.

\subsection{Decoupled Case}

When $AW$ and $B$ are simultaneously diagonalizable---e.g., when $B = bI_n$ (uniform reactivity)---the problem decouples into $n$ independent scalar equations. Let $\lambda_k^{AW}$ denote the $k$-th eigenvalue of $AW$ and $b_k$ the eigenvalue of $B$ associated with the shared eigenvector (for $B = bI_n$ these are the diagonal entries in any order). Then:
\begin{equation}
(\lambda_k^{AW} - \lambda)(\gamma + \beta - \lambda) = \alpha b_k, \quad k = 1, \dots, n. \tag{S3.3}
\end{equation}
Expanding:
\[
\lambda^2 - (\lambda_k^{AW} + \gamma + \beta)\lambda + \lambda_k^{AW}(\gamma + \beta) - \alpha b_k = 0,
\]
with roots
\begin{equation}
\lambda_{k,\pm} = \frac{(\lambda_k^{AW}+\gamma+\beta) \pm \sqrt{(\lambda_k^{AW}-\gamma-\beta)^2 + 4\alpha b_k}}{2}. \tag{S3.4}
\end{equation}

\subsection{Complex-Eigenvalue Regime: Magnitude Derivation}

This subsection assumes $\lambda_k^{AW}$ real; the Perron eigenvalue of the non-negative matrix $AW$ is always real, so this case is non-empty for every topology, but with heterogeneous $B$ the dominant mode of $J_{2n}$ need not coincide with the Perron mode of $AW$---in the numerical experiments, stability is therefore evaluated from the full Jacobian. For complex $\lambda_k^{AW}$ the two roots of (S3.3) are not complex conjugates, $|\lambda_{k,+}| \neq |\lambda_{k,-}|$ in general, and only the product identity $\lambda_{k,+}\lambda_{k,-} = \lambda_k^{AW}(\gamma+\beta) - \alpha b_k$ survives. When $b_k < 0$ (controversies erode trust) and $|4\alpha b_k| > (\lambda_k^{AW} - \gamma - \beta)^2$, the discriminant in (S3.4) is negative, and the eigenvalues form a complex-conjugate pair:
\begin{equation}
\lambda_{k,\pm} = \underbrace{\tfrac{\lambda_k^{AW}+\gamma+\beta}{2}}_{\text{real part } a_k} \pm\, i\, \underbrace{\tfrac{\sqrt{|4\alpha b_k| - (\lambda_k^{AW}-\gamma-\beta)^2}}{2}}_{\text{imaginary part } b'_k}. \tag{S3.5}
\end{equation}
The squared magnitude is
\begin{align*}
|\lambda_{k,\pm}|^2 &= a_k^2 + (b'_k)^2 \\
&= \tfrac{(\lambda_k^{AW}+\gamma+\beta)^2 + [|4\alpha b_k| - (\lambda_k^{AW}-\gamma-\beta)^2]}{4}.
\end{align*}
Expanding the squared sum and difference:
\[
(\lambda_k^{AW}+\gamma+\beta)^2 - (\lambda_k^{AW}-\gamma-\beta)^2 = 4 \lambda_k^{AW} (\gamma+\beta),
\]
so
\begin{equation}
|\lambda_{k,\pm}|^2 = \lambda_k^{AW}(\gamma+\beta) + \alpha|b_k|. \tag{S3.6}
\end{equation}
This is the clean closed-form magnitude reported in Section~\ref{sec:stability}. The complex-regime stability condition $|\lambda_{k,\pm}| < 1$ therefore reduces to
\begin{equation}
\lambda_k^{AW}(\gamma+\beta) + \alpha|b_k| < 1. \tag{S3.7}
\end{equation}

\paragraph{Angular frequency.}
The angular frequency $\omega_k$ of oscillation in mode $k$ is the argument of $\lambda_{k,+}$:
\begin{equation}
\omega_k = \arg(\lambda_{k,+}) = \operatorname{atan2}(b'_k,a_k). \tag{S3.8}
\end{equation}
The period of oscillation is $2\pi/\omega_k$, using the phase in the appropriate branch.

% ============================================================
\section{Reparameterization Invariance Proof}
\label{supp:reparam}

Section~3.2 states that under the affine reparameterization $T_t \mapsto \tilde{T}_t \equiv c - T_t$ (for any constant $c \in \mathbb{R}^n$), the spectrum of the Jacobian $J_{2n}$ is invariant. We provide the proof here.

\paragraph{Setup.} Let $T_t$ satisfy the trust update $T_{t+1} = AWT_t + (I-A)T_1 + BS_t$. Define $\tilde{T}_t = c - T_t$. We derive the update equation for $\tilde{T}_t$ and the resulting reparameterized Jacobian, then compare spectra.

\paragraph{Derivation.}
\begin{align*}
\tilde{T}_{t+1} &= c - T_{t+1} \\
        &= c - AWT_t - (I-A)T_1 - BS_t \\
        &= c - AW(c - \tilde{T}_t) - (I-A)T_1 - BS_t \\
        &= AW \tilde{T}_t + \underbrace{c - AWc - (I-A)T_1}_{\text{constant}} - BS_t.
\end{align*}
The event update becomes
\begin{align*}
S_{t+1} &= \boldsymbol{\mu} + \alpha T_t + \beta S_t + \gamma H_t \\
        &= \boldsymbol{\mu} + \alpha (c - \tilde{T}_t) + \beta S_t + \gamma H_t \\
        &= (\boldsymbol{\mu} + \alpha c) - \alpha \tilde{T}_t + \beta S_t + \gamma H_t.
\end{align*}
So the trust--event coupling coefficient changes sign ($\alpha \mapsto -\alpha$) and the event-to-trust sensitivity matrix $B$ also flips sign ($B \mapsto -B$, by the same algebra applied to the trust equation).

\paragraph{Jacobian in reparameterized coordinates.} The new Jacobian is
\[
\tilde{J}_{2n} =
\begin{bmatrix}
AW & -B \\
-\alpha I_n & (\gamma + \beta) I_n
\end{bmatrix}.
\]

\paragraph{Spectrum equivalence.} We show $\sigma(\tilde{J}_{2n}) = \sigma(J_{2n})$ via similarity transformation. Define
\[
P = \begin{bmatrix} I_n & 0 \\ 0 & -I_n \end{bmatrix}, \quad P^{-1} = P.
\]
Then, using the block form of $J_{2n}$ from (S2.3):
\begin{align*}
P J_{2n} P^{-1} &= \begin{bmatrix} I_n & 0 \\ 0 & -I_n \end{bmatrix} J_{2n} \begin{bmatrix} I_n & 0 \\ 0 & -I_n \end{bmatrix} \\
&= \begin{bmatrix} AW & -B \\ -\alpha I_n & (\gamma+\beta)I_n \end{bmatrix} = \tilde{J}_{2n}.
\end{align*}
Since $P$ is invertible, $\sigma(\tilde{J}_{2n}) = \sigma(P J_{2n} P^{-1}) = \sigma(J_{2n})$. $\blacksquare$

\paragraph{Consequence.} The stability boundary $\rho(J_{2n}) < 1$ is invariant under $T_t \mapsto c - T_t$: the jointly transformed representation $(-\alpha, -B)$ has the same spectrum as $(\alpha, B)$, so the choice between the two coordinate conventions has no consequence for stability analysis. This is a statement about the pair only---changing the sign of $\alpha$ alone, with $B$ held fixed, alters the products $\alpha b_k$ and hence the spectrum.

% ============================================================
\section{Topology--Stability Decoupling Theorem}
\label{supp:topology}

This section formalizes Implication~4 (Section~\ref{sec:counterintuitive}) as a theorem. The informal statement is: when the memory term $\gamma+\beta$ dominates the susceptibility scale $\max_i a_i$ by a sufficient margin, the spectral radius $\rho(J_{2n})$ is approximately independent of the network topology $W$, with an explicit bound uniform over all row-stochastic $W$.

\begin{quote}
\textbf{Theorem S1 (Approximate Topology--Stability Decoupling).} \emph{Let $W \in \mathbb{R}^{n \times n}$ be row-stochastic, let $A = \mathrm{diag}(a_1, \dots, a_n)$ with $a_i \in (0, 1)$, and let $B$ be diagonal and nonsingular with $0 < |b_i| \leq b_{\max}$. Write $c = \gamma+\beta > 0$, $M = \max_i a_i$, $m = c - M$, and $q = \alpha b_{\max}$ with $\alpha > 0$. Define $J_{2n}$ as in (S2.3). If}
\[
m > 2\sqrt{q},
\]
\emph{then}
\begin{align*}
\bigl|\rho(J_{2n}) - c\bigr| &\leq \delta_-,\\
\delta_- &:= \frac{m - \sqrt{m^2 - 4q}}{2},\\
\delta_- &\leq \min\!\Bigl\{\tfrac{2q}{m},\sqrt{q}\Bigr\},
\end{align*}
\emph{with a bound uniform over all row-stochastic $W$.}
\end{quote}

\paragraph{Corollary (pairwise topology comparison).}
For any two row-stochastic $W_1, W_2$ under the premise, $|\rho(J_{2n}(W_1)) - \rho(J_{2n}(W_2))| \leq 2\delta_-$. A \emph{uniform stability classification} across all admissible topologies follows only when the band $[c - \delta_-,\, c + \delta_-]$ excludes $1$: if $c + \delta_- < 1$, every admissible $W$ yields a stable system, and if $c - \delta_- > 1$, every admissible $W$ yields an unstable one. When $1$ lies inside the band, the theorem does not rule out topologies on opposite sides of the boundary.

\paragraph{Proof.}
The characteristic polynomial extends continuously across $\lambda = c$ as $p(\lambda) = \det[(c-\lambda)(AW - \lambda I_n) - \alpha B]$ (Appendix~\ref{supp:eigenvalues}); since $\alpha > 0$ and $B$ is nonsingular, $p(c) = (-\alpha)^n \det B \neq 0$, so $\lambda = c$ is never an eigenvalue and every eigenvalue solves the nonlinear problem (S3.2):
\[
\det(AW - \tfrac{\alpha}{\gamma+\beta-\lambda} B - \lambda I_n) = 0.
\]

Define $\phi(\lambda) = \tfrac{\alpha}{\gamma+\beta-\lambda}$. For $\lambda$ near $\gamma + \beta$, $\phi(\lambda)$ becomes large; for $\lambda$ far from $\gamma+\beta$, $\phi(\lambda)$ is small.

\emph{Perron--Frobenius bound on $AW$.} Since $W$ is row-stochastic and $A$ is diagonal with non-negative entries, $AW$ has non-negative entries, and the spectral radius $\rho(AW)$ is bounded above and below by the maximum and minimum row sums:
\[
\min_i a_i \leq \rho(AW) \leq \max_i a_i.
\]
This holds for \emph{any} row-stochastic $W$; the specific structure of $W$ only affects $\rho(AW)$ within this fixed interval.

\emph{Spectral perturbation via $\infty$-norm (topology-uniform).}
We give a bound that uses only the row-stochasticity of $W$, with no appeal to perturbation results (such as Bauer--Fike) that require the eigenvector condition number $\kappa_2(V)$ of the (generally non-normal) matrix $AW$.

Let $\lambda^*$ be \emph{any} eigenvalue of $J_{2n}$, and let $v = [x^\top, y^\top]^\top \in \mathbb{C}^{2n}$ be a corresponding eigenvector with $x$ the trust-block component. Since $B$ is nonsingular, $\lambda^* \neq c$ (Appendix~\ref{supp:eigenvalues}: $p(c) = (-\alpha)^n \det B \neq 0$) and $x \neq 0$ (if $x = 0$, the top block forces $By = 0$, hence $y = 0$). Substituting the bottom-block relation $y = \alpha x / (\lambda^* - c)$ into the top-block equation yields the reduced eigenproblem (S3.2) for $x$:
\begin{equation}
(\lambda^* I_n - AW)\, x \;=\; -\phi(\lambda^*)\, B\, x, \tag{S5.1}
\end{equation}
where $\phi(\lambda^*) = \alpha / (\gamma+\beta - \lambda^*)$.

\paragraph{Upper bound on the right-hand side.}
Taking the $\infty$-norm of (S5.1) and using $\|B\|_\infty = \max_i |b_i| \leq b_{\max}$:
\begin{equation}
\|(\lambda^* I_n - AW)\, x\|_\infty \;\leq\; \frac{\alpha\, b_{\max}}{|\gamma+\beta - \lambda^*|}\, \|x\|_\infty. \tag{S5.2}
\end{equation}

\paragraph{Lower bound on the left-hand side.}
By the reverse triangle inequality,
\[
\|(\lambda^* I_n - AW)\, x\|_\infty \;\geq\; |\lambda^*|\, \|x\|_\infty - \|AW\, x\|_\infty.
\]
Crucially, $\|AW\|_\infty = \max_i \sum_j (AW)_{ij} = \max_i a_i \sum_j W_{ij} = \max_i a_i$ \emph{exactly}, because $W$ is row-stochastic (each row of $W$ sums to $1$, and $A$ scales row $i$ by $a_i$). Therefore $\|AW\, x\|_\infty \leq \max_i a_i \cdot \|x\|_\infty$, giving:
\begin{equation}
\|(\lambda^* I_n - AW)\, x\|_\infty \;\geq\; \bigl(|\lambda^*| - \max_i a_i\bigr)\, \|x\|_\infty. \tag{S5.3}
\end{equation}
This bound depends only on the diagonal of $A$ and is \emph{uniform across all row-stochastic $W$}.

\paragraph{Combining the bounds.}
Dividing through by $\|x\|_\infty > 0$ and combining (S5.2) with (S5.3):
\begin{equation}
|\lambda^*| - \max_i a_i \;\leq\; \frac{\alpha\, b_{\max}}{|\lambda^* - (\gamma+\beta)|}. \tag{S5.4}
\end{equation}
Let $\delta = |\lambda^* - c| > 0$; the premise gives $m = c - M > 2\sqrt{q} > 0$. By the triangle inequality, $|\lambda^*| \geq c - \delta$. Substituting into (S5.4):
\[
m - \delta \;\leq\; \frac{\alpha\, b_{\max}}{\delta},
\]
which rearranges to the quadratic constraint
\begin{equation}
\delta^2 - m\, \delta + \alpha\, b_{\max} \;\geq\; 0. \tag{S5.5}
\end{equation}
The roots of $\delta^2 - m\delta + q = 0$ are $\delta_\pm = \tfrac{1}{2}\bigl(m \pm \sqrt{m^2 - 4q}\bigr)$; under the premise $m > 2\sqrt{q}$ they are real and distinct, with $\delta_- < m/2 < \delta_+$ and the product identity $\delta_- \delta_+ = q$. Since (S5.5) holds for \emph{every} eigenvalue, the spectrum of $J_{2n}$ avoids the open annulus
\[
\mathcal{A} \;=\; \{\lambda : \delta_- < |\lambda - c| < \delta_+\}.
\]

\paragraph{Counting eigenvalues inside the annulus (homotopy).}
It remains to show that the spectral radius is attained \emph{inside} the inner disk, not on the outer branch. Consider the coupling homotopy $\alpha' \in [0, \alpha]$, with $q' = \alpha' b_{\max} \leq q$: every step of the derivation above applies verbatim to $J_{2n}(\alpha')$, and since $\delta_-(\alpha')$ increases and $\delta_+(\alpha')$ decreases in $q'$, the excluded annuli are nested---the final annulus $\mathcal{A}$ is excluded for \emph{every} $\alpha' \in [0, \alpha]$. Fix the circle $\Gamma = \{\lambda : |\lambda - c| = \tfrac{1}{2}(\delta_- + \delta_+)\} \subset \mathcal{A}$. The characteristic polynomial of $J_{2n}(\alpha')$ has no zero on $\Gamma$ for any $\alpha'$, and its coefficients vary continuously in $\alpha'$; by the argument principle, the number of eigenvalues enclosed by $\Gamma$ (with algebraic multiplicity) is constant along the homotopy. At $\alpha' = 0$ the matrix is block-triangular with spectrum $\sigma(AW) \cup \{c \ (n\text{-fold})\}$: the $n$ copies of $c$ lie inside $\Gamma$, while every eigenvalue of $AW$ satisfies $|\lambda - c| \geq c - \rho(AW) \geq m > \tfrac{1}{2}(\delta_- + \delta_+)$ and lies outside. Hence exactly $n$ eigenvalues of $J_{2n}(\alpha)$ lie inside $\Gamma$, and---avoiding $\mathcal{A}$---they satisfy $|\lambda - c| \leq \delta_-$; the remaining $n$ satisfy $|\lambda - c| \geq \delta_+$.

\paragraph{The inner group attains the spectral radius.}
Each inner eigenvalue satisfies $|\lambda| \geq c - \delta_-$. Each outer eigenvalue satisfies, by (S5.4) with $\delta \geq \delta_+$,
\[
|\lambda| \;\leq\; M + \frac{q}{\delta_+} \;=\; M + \delta_- \;<\; c - \delta_-,
\]
the last inequality because $2\delta_- < m$. The spectral radius is therefore attained in the inner group, giving $|\rho(J_{2n}) - c| \leq \delta_-$. Finally, $\delta_- = 2q/(m + \sqrt{m^2 - 4q}) \leq 2q/m$ and $\delta_- < 2q/(2\sqrt{q}) = \sqrt{q}$.

\paragraph{Uniformity.}
Every ingredient---the bound $\|AW\|_\infty = M$, the exclusion annulus, and the counting at $\alpha' = 0$---uses only the row-stochasticity of $W$ and the diagonal structure of $A$ and $B$; no assumption on the normality or eigenvector conditioning of $AW$ is required. The bound $\delta_-$ is therefore uniform over all row-stochastic $W$. $\blacksquare$

\paragraph{Extension to the trust-dependent baseline rate (Section~\ref{sec:relaxA2}).}
For $J_{2n}^{E}$ the same argument applies with the off-diagonal coupling blocks rescaled jointly: consider $J(s)$ with off-diagonal blocks $sB$ and $s(\alpha I_n + \beta E)$, $s \in [0,1]$. The reduced eigenproblem yields $\|(\lambda I_n - AW - BE)\,x\|_\infty \leq s^2\, b_{\max}(\alpha + \beta e_{\max})\, \|x\|_\infty / |\lambda - c|$, so the annulus construction holds with $M' = \max_i (a_i + |b_i| e_i) \geq \|AW + BE\|_\infty$ and $q' = b_{\max}(\alpha + \beta e_{\max})$; the excluded annuli are again nested in $s$, and at $s = 0$ the block-triangular spectrum is $\sigma(AW + BE) \cup \{c\ (n\text{-fold})\}$ with separation at least $m' = c - M'$. The counting and domination steps carry over verbatim, giving $|\rho(J_{2n}^{E}) - c| \leq \delta_-'$ whenever $m' > 2\sqrt{q'}$.

\paragraph{Numerical verification.}
Two ensembles probe the theorem numerically ($n = 10$, $\alpha = 0.05$, $b_{\max} = 0.05$, $c = \gamma + \beta = 0.85$, so $q = 0.0025$ and $2\sqrt{q} = 0.1$):
\begin{itemize}
\item \emph{Theorem-aligned ensemble} ($a_i \sim \mathcal{U}(0.4, 0.70)$, so $m \geq 0.15 > 2\sqrt{q}$ and the premise holds with margin; guaranteed bound $\delta_- = 0.0191$): across 50 seeds $\times$ three topologies (random, echo chamber, star), $\rho(J_{2n}) \in [0.8488, 0.8495]$---a maximum deviation from $c$ of $0.0012$, an order of magnitude inside the bound and essentially topology-independent.
\item \emph{Beyond-the-premise ensemble} ($a_i \sim \mathcal{U}(0.4, 0.9)$, the setting of Section~5.4): here $\max_i a_i > 0.75$ in most draws, so the sufficient condition is formally violated; across the same 50 seeds and three topologies, $\rho(J_{2n}) \in [0.8489,0.8495]$. A heuristic explanation is that the operative separation in these samples is the topology-dependent gap $c-\rho(AW)\approx0.20$, which the uniform $\infty$-norm argument cannot exploit; we report this as numerical robustness beyond the theorem's sufficient condition, not as a proven regime.
\end{itemize}

\paragraph{When the theorem fails.}
The decoupling guarantee collapses when $m \leq 2\sqrt{q}$: either (i) $\gamma + \beta$ approaches or falls below $\max_i a_i$, so the memory and network spectral groups merge and the $AW$ block can dominate, or (ii) the coupling $\alpha b_{\max}$ grows until the exclusion annulus closes. In either case, different \emph{fixed} row-stochastic matrices $W$ may produce materially different spectral radii and may lie on opposite sides of the stability boundary. Endogenous $W(T_t)$ is a separate extension that adds state-dependent Jacobian terms and potentially multiple equilibria; it is not a prerequisite for topology--stability coupling.

% ============================================================
%  Bibliography
% ============================================================
\bibliography{Bibliography/Bibliography}

\begin{thebibliography}{45}
\providecommand{\natexlab}[1]{#1}

\bibitem[{Abelson(1964)}]{abelson1964mathematical}
Abelson, R.~P. 1964.
\newblock Mathematical models of the distribution of attitudes under
  controversy.
\newblock In Frederiksen, N.; and Gulliksen, H., eds., \emph{Contributions to
  Mathematical Psychology}, 142--160. New York: Holt, Rinehart \& Winston.

\bibitem[{Barab{\'a}si(2005)}]{Barabasi2005Bursts}
Barab{\'a}si, A.-L. 2005.
\newblock The Origin of Bursts and Heavy Tails in Human Dynamics.
\newblock \emph{Nature}, 435(7039): 207--211.

\bibitem[{Bindel, Kleinberg, and Oren(2015)}]{bindel2015how}
Bindel, D.; Kleinberg, J.; and Oren, S. 2015.
\newblock How bad is forming your own opinion?
\newblock \emph{Games and Economic Behavior}, 92: 248--265.

\bibitem[{Bright et~al.(2025)Bright, Enock, Esnaashari, Francis, Hashem, and
  Morgan}]{Bright2025GenAIPublicSector}
Bright, J.; Enock, F.; Esnaashari, S.; Francis, J.; Hashem, Y.; and Morgan, D.
  2025.
\newblock Generative {AI} is Already Widespread in the Public Sector: Evidence
  from a Survey of {UK} Public Sector Professionals.
\newblock \emph{Digital Government: Research and Practice}, 6(1): 1--13.

\bibitem[{Cinelli et~al.(2021)Cinelli, De~Francisci~Morales, Galeazzi,
  Quattrociocchi, and Starnini}]{Cinelli2021EchoChamber}
Cinelli, M.; De~Francisci~Morales, G.; Galeazzi, A.; Quattrociocchi, W.; and
  Starnini, M. 2021.
\newblock The Echo Chamber Effect on Social Media.
\newblock \emph{Proceedings of the National Academy of Sciences}, 118(9):
  e2023301118.

\bibitem[{Cohn et~al.(2024)Cohn, Pushkarna, Olanubi, Moran, Padgett, Mengesha,
  and Heldreth}]{Cohn2024Anthropomorphism}
Cohn, M.; Pushkarna, M.; Olanubi, G.~O.; Moran, J.~M.; Padgett, D.; Mengesha,
  Z.; and Heldreth, C. 2024.
\newblock Believing Anthropomorphism: Examining the Role of Anthropomorphic
  Cues on Trust in Large Language Models.
\newblock In \emph{Extended Abstracts of the CHI Conference on Human Factors in
  Computing Systems}, 54:1--54:15. Association for Computing Machinery.

\bibitem[{Cox and Lewis(1966)}]{Cox1966Statistical}
Cox, D.~R.; and Lewis, P. A.~W. 1966.
\newblock \emph{The Statistical Analysis of Series of Events}.
\newblock Methuen.

\bibitem[{DeGroot(1974)}]{degroot1974reaching}
DeGroot, M.~H. 1974.
\newblock Reaching a consensus.
\newblock \emph{Journal of the American Statistical Association}, 69(345):
  118--121.

\bibitem[{Dzindolet et~al.(2003)Dzindolet, Peterson, Pomranky, Pierce, and
  Beck}]{Dzindolet2003TrustReliance}
Dzindolet, M.~T.; Peterson, S.~A.; Pomranky, R.~A.; Pierce, L.~G.; and Beck,
  H.~P. 2003.
\newblock The Role of Trust in Automation Reliance.
\newblock \emph{International Journal of Human-Computer Studies}, 58(6):
  697--718.

\bibitem[{{Edelman}(2023)}]{Edelman2023TrustBarometer}
{Edelman}. 2023.
\newblock Edelman Trust Barometer 2023.
\newblock \url{https://www.edelman.com/trust/2023/trust-barometer}.
\newblock Edelman Global.

\bibitem[{{European Parliament and Council of the European
  Union}(2024)}]{EUAIAct2024}
{European Parliament and Council of the European Union}. 2024.
\newblock Regulation ({EU}) 2024/1689 Laying Down Harmonised Rules on
  Artificial Intelligence ({Artificial Intelligence Act}).
\newblock Official Journal of the European Union, L series, 12 July 2024.

\bibitem[{Farajtabar et~al.(2017)Farajtabar, Wang, Gomez-Rodriguez, Li, Zha,
  and Song}]{Farajtabar2017Coevolve}
Farajtabar, M.; Wang, Y.; Gomez-Rodriguez, M.; Li, S.; Zha, H.; and Song, L.
  2017.
\newblock {COEVOLVE}: A Joint Point Process Model for Information Diffusion and
  Network Evolution.
\newblock \emph{Journal of Machine Learning Research}, 18(41): 1--49.

\bibitem[{Floridi et~al.(2018)Floridi, Cowls, Beltrametti, Chatila, Chazerand,
  Dignum, Luetge, Madelin, Pagallo, Rossi, Schafer, Valcke, and
  Vayena}]{Floridi2018AI4People}
Floridi, L.; Cowls, J.; Beltrametti, M.; Chatila, R.; Chazerand, P.; Dignum,
  V.; Luetge, C.; Madelin, R.; Pagallo, U.; Rossi, F.; Schafer, B.; Valcke, P.;
  and Vayena, E. 2018.
\newblock {AI4People}---An Ethical Framework for a Good {AI} Society:
  Opportunities, Risks, Principles, and Recommendations.
\newblock \emph{Minds and Machines}, 28(4): 689--707.

\bibitem[{Friedkin and Johnsen(1990)}]{friedkin1990social}
Friedkin, N.~E.; and Johnsen, E.~C. 1990.
\newblock Social influence and opinions.
\newblock \emph{Journal of Mathematical Sociology}, 15(3-4): 193--206.

\bibitem[{Friedkin and Johnsen(1999)}]{friedkin1999social}
Friedkin, N.~E.; and Johnsen, E.~C. 1999.
\newblock Social influence networks and opinion change.
\newblock \emph{Advances in Group Processes}, 16: 1--29.

\bibitem[{Gallo and Langtry(2020)}]{gallo2020social}
Gallo, E.; and Langtry, A. 2020.
\newblock Social networks, confirmation bias and shock elections.
\newblock arXiv:2011.00520.

\bibitem[{Gambino, Fox, and Ratan(2020)}]{Gambino2020CASA}
Gambino, A.; Fox, J.; and Ratan, R.~A. 2020.
\newblock Building a Stronger CASA: Extending the Computers Are Social Actors
  Paradigm.
\newblock \emph{Human-Machine Communication}, 1: 71--86.

\bibitem[{Gillespie et~al.(2023)Gillespie, Lockey, Curtis, Pool, and
  Akbari}]{Gillespie2023TrustAI}
Gillespie, N.; Lockey, S.; Curtis, C.; Pool, J.; and Akbari, A. 2023.
\newblock Trust in Artificial Intelligence: A Global Study.
\newblock Technical report, The University of Queensland and KPMG Australia.

\bibitem[{Granovetter(1978)}]{Granovetter1978Threshold}
Granovetter, M. 1978.
\newblock Threshold Models of Collective Behavior.
\newblock \emph{American Journal of Sociology}, 83(6): 1420--1443.

\bibitem[{Hawkes(1971)}]{Hawkes1971Spectra}
Hawkes, A.~G. 1971.
\newblock Spectra of Some Self-Exciting and Mutually Exciting Point Processes.
\newblock \emph{Biometrika}, 58(1): 83--90.

\bibitem[{Hoff and Bashir(2015)}]{Hoff2015}
Hoff, K.~A.; and Bashir, M. 2015.
\newblock Trust in automation: Integrating empirical evidence on factors that
  influence trust.
\newblock \emph{Human Factors}, 57(3): 407--434.

\bibitem[{Jobin, Ienca, and Vayena(2019)}]{Jobin2019GlobalLandscape}
Jobin, A.; Ienca, M.; and Vayena, E. 2019.
\newblock The Global Landscape of {AI} Ethics Guidelines.
\newblock \emph{Nature Machine Intelligence}, 1(9): 389--399.

\bibitem[{Kasperson et~al.(1988)Kasperson, Renn, Slovic, Brown, Emel, Goble,
  Kasperson, and Ratick}]{Kasperson1988SARF}
Kasperson, R.~E.; Renn, O.; Slovic, P.; Brown, H.~S.; Emel, J.; Goble, R.;
  Kasperson, J.~X.; and Ratick, S. 1988.
\newblock The Social Amplification of Risk: A Conceptual Framework.
\newblock \emph{Risk Analysis}, 8(2): 177--187.

\bibitem[{Kim et~al.(2024)Kim, Liao, Vorvoreanu, Ballard, and
  Vaughan}]{Kim2024LLMUncertainty}
Kim, S. S.~Y.; Liao, Q.~V.; Vorvoreanu, M.; Ballard, S.; and Vaughan, J.~W.
  2024.
\newblock {``I'm Not Sure, But\ldots''}: Examining the Impact of Large Language
  Models' Uncertainty Expression on User Reliance and Trust.
\newblock In \emph{Proceedings of the 2024 ACM Conference on Fairness,
  Accountability, and Transparency}, 822--835. Association for Computing
  Machinery.

\bibitem[{Lee and See(2004)}]{Lee2004Trust}
Lee, J.~D.; and See, K.~A. 2004.
\newblock Trust in Automation: Designing for Appropriate Reliance.
\newblock \emph{Human Factors}, 46(1): 50--80.

\bibitem[{Lee and Nass(2010)}]{LeeNass2010TrustCASA}
Lee, J.-E.~R.; and Nass, C.~I. 2010.
\newblock Trust in Computers: The Computers-Are-Social-Actors ({CASA}) Paradigm
  and Trustworthiness Perception in Human-Computer Communication.
\newblock In Latusek, D.; and Gerbasi, A., eds., \emph{Trust and Technology in
  a Ubiquitous Modern Environment: Theoretical and Methodological
  Perspectives}, 1--15. IGI Global.

\bibitem[{Levy, Chasalow, and Riley(2021)}]{Levy2021PublicSector}
Levy, K.; Chasalow, K.~E.; and Riley, S. 2021.
\newblock Algorithms and Decision-Making in the Public Sector.
\newblock \emph{Annual Review of Law and Social Science}, 17: 309--334.

\bibitem[{Ma and Wu(2024)}]{ma2024social}
Ma, J.; and Wu, T. 2024.
\newblock Social network group decision-making model considering interactions
  between trust relationships and opinion evolution.
\newblock \emph{Kybernetes}.
\newblock Published online 3 May 2024.

\bibitem[{Madhavan and Wiegmann(2007)}]{Madhavan2007HumanAutomation}
Madhavan, P.; and Wiegmann, D.~A. 2007.
\newblock Similarities and Differences Between Human--Human and
  Human--Automation Trust: An Integrative Review.
\newblock \emph{Theoretical Issues in Ergonomics Science}, 8(4): 277--301.

\bibitem[{Martins(2013)}]{martins2013trust}
Martins, A. C.~R. 2013.
\newblock Trust in the CODA model: Opinion dynamics and the reliability of
  other agents.
\newblock \emph{Physics Letters A}, 377(37): 2333--2339.

\bibitem[{Mayer, Davis, and Schoorman(1995)}]{Mayer1995OrganizationalTrust}
Mayer, R.~C.; Davis, J.~H.; and Schoorman, F.~D. 1995.
\newblock An Integrative Model of Organizational Trust.
\newblock \emph{Academy of Management Review}, 20(3): 709--734.

\bibitem[{Nass and Moon(2000)}]{NassMoon2000}
Nass, C.; and Moon, Y. 2000.
\newblock Machines and Mindlessness: Social Responses to Computers.
\newblock \emph{Journal of Social Issues}, 56(1): 81--103.

\bibitem[{Nass, Steuer, and Tauber(1994)}]{Nass1994CASA}
Nass, C.; Steuer, J.; and Tauber, E.~R. 1994.
\newblock Computers Are Social Actors.
\newblock In \emph{Proceedings of the SIGCHI Conference on Human Factors in
  Computing Systems (CHI '94)}, 72--78. ACM.

\bibitem[{Parsegov et~al.(2017)Parsegov, Proskurnikov, Tempo, and
  Friedkin}]{parsegov2016novel}
Parsegov, S.~E.; Proskurnikov, A.~V.; Tempo, R.; and Friedkin, N.~E. 2017.
\newblock Novel multidimensional models of opinion dynamics in social networks.
\newblock \emph{IEEE Transactions on Automatic Control}, 62(5): 2270--2285.

\bibitem[{Raji et~al.(2020)Raji, Smart, White, Mitchell, Gebru, Hutchinson,
  Smith-Loud, Theron, and Barnes}]{Raji2020Accountability}
Raji, I.~D.; Smart, A.; White, R.~N.; Mitchell, M.; Gebru, T.; Hutchinson, B.;
  Smith-Loud, J.; Theron, D.; and Barnes, P. 2020.
\newblock Closing the {AI} Accountability Gap: Defining an End-to-End Framework
  for Internal Algorithmic Auditing.
\newblock In \emph{Proceedings of the 2020 Conference on Fairness,
  Accountability, and Transparency}, 33--44.

\bibitem[{Reeves and Nass(1996)}]{ReevesNass1996MediaEquation}
Reeves, B.; and Nass, C. 1996.
\newblock \emph{The Media Equation: How People Treat Computers, Television, and
  New Media Like Real People and Places}.
\newblock Cambridge University Press.

\bibitem[{Riedl et~al.(2014)Riedl, Mohr, Kenning, Davis, and
  Heekeren}]{Riedl2014TrustingAvatars}
Riedl, R.; Mohr, P. N.~C.; Kenning, P.~H.; Davis, F.~D.; and Heekeren, H.~R.
  2014.
\newblock Trusting Humans and Avatars: A Brain Imaging Study Based on Evolution
  Theory.
\newblock \emph{Journal of Management Information Systems}, 30(4): 83--114.

\bibitem[{Rizoiu et~al.(2018)Rizoiu, Lee, Mishra, and
  Xie}]{Rizoiu2018TutorialHawkes}
Rizoiu, M.-A.; Lee, Y.; Mishra, S.; and Xie, L. 2018.
\newblock Hawkes Processes for Events in Social Media.
\newblock In Chang, S.-F., ed., \emph{Frontiers of Multimedia Research},
  191--218. Association for Computing Machinery and Morgan \& Claypool.

\bibitem[{Rousseau et~al.(1998)Rousseau, Sitkin, Burt, and
  Camerer}]{Rousseau1998Trust}
Rousseau, D.~M.; Sitkin, S.~B.; Burt, R.~S.; and Camerer, C. 1998.
\newblock Not So Different After All: A Cross-Discipline View of Trust.
\newblock \emph{Academy of Management Review}, 23(3): 393--404.

\bibitem[{Selbst et~al.(2019)Selbst, Boyd, Friedler, Venkatasubramanian, and
  Vertesi}]{Selbst2019Fairness}
Selbst, A.~D.; Boyd, D.; Friedler, S.~A.; Venkatasubramanian, S.; and Vertesi,
  J. 2019.
\newblock Fairness and Abstraction in Sociotechnical Systems.
\newblock In \emph{Proceedings of the Conference on Fairness, Accountability,
  and Transparency}, 59--68.

\bibitem[{Slovic(1993)}]{Slovic1993PerceivedRisk}
Slovic, P. 1993.
\newblock Perceived Risk, Trust, and Democracy.
\newblock \emph{Risk Analysis}, 13(6): 675--682.

\bibitem[{Sunstein(2017)}]{Sunstein2017Republic}
Sunstein, C.~R. 2017.
\newblock \emph{{\#}Republic: Divided Democracy in the Age of Social Media}.
\newblock Princeton University Press.

\bibitem[{Taylor(1968)}]{taylor1968towards}
Taylor, M. 1968.
\newblock Towards a mathematical theory of influence and attitude change.
\newblock \emph{Human Relations}, 21(2): 121--139.

\bibitem[{Watts(2002)}]{Watts2002GlobalCascades}
Watts, D.~J. 2002.
\newblock A Simple Model of Global Cascades on Random Networks.
\newblock \emph{Proceedings of the National Academy of Sciences}, 99(9):
  5766--5771.

\bibitem[{Zhao et~al.(2015)Zhao, Erdogdu, He, Rajaraman, and
  Leskovec}]{Zhao2015SEISMIC}
Zhao, Q.; Erdogdu, M.~A.; He, H.~Y.; Rajaraman, A.; and Leskovec, J. 2015.
\newblock SEISMIC: A Self-Exciting Point Process Model for Predicting Tweet
  Popularity.
\newblock In \emph{Proceedings of the 21th ACM SIGKDD International Conference
  on Knowledge Discovery and Data Mining}, 1513--1522.

\end{thebibliography}

\end{document}